\documentclass[aps,preprint,showpacs,groupedaddress,floatfix]{revtex4}
\usepackage{graphicx}
\usepackage{dcolumn}
\usepackage{bm}

\newcommand{\lp}{\left}
\newcommand{\rp}{\right}

\begin{document}

\title{Quasiparticle random phase approximation based on the relativistic
Hartree-Bogoliubov model II:\\ Nuclear spin and isospin excitations}
\author{N. Paar}
\affiliation{
Institut f\" ur Kernphysik, Technische Universit\" at Darmstadt 
Schlossgartenstr. 9,
D-64289 Darmstadt, Germany}
\author{T. Nik\v si\' c}
\author{D. Vretenar}
\affiliation{Physics Department, Faculty of Science, University of Zagreb, 
Croatia, and \\
Physik-Department der Technischen Universit\"at M\"unchen, D-85748 Garching,
Germany}
\author{P. Ring}
\affiliation{Physik-Department der Technischen Universit\"at M\"unchen, 
D-85748 Garching,
Germany}
\date{\today}

\begin{abstract}
The proton-neutron relativistic quasiparticle random 
phase approximation (PN-RQRPA) is
formulated in the canonical single-nucleon 
basis of the relativistic Hartree-Bogoliubov (RHB) model, 
for an effective Lagrangian characterized by 
density-dependent meson-nucleon couplings.
The model includes both 
the $T=1$ and $T=0$ pairing channels.
Pair configurations formed from the fully or 
partially occupied states of
positive energy in the Fermi sea, 
and the empty negative-energy states from the Dirac sea, are
included in PN-RQRPA configuration space.
The model is applied to the analysis of charge-exchange modes:
isobaric analog resonances and  Gamow-Teller resonances.
\end{abstract}

\pacs{21.30.Fe, 21.60.Jz, 24.30.Cz, 25.40.Kv}
\maketitle

\bigskip \bigskip

\section{\label{secI}Introduction}

A consistent and unified treatment of mean-field and
pairing correlations is crucial for a quantitative analysis 
of ground-state properties and multipole response of unstable,
weakly bound nuclei far from the line of $\beta$-stability.
In Ref.~\cite{Paa.03} we have formulated 
the relativistic quasiparticle random phase approximation (RQRPA) 
in the canonical single-nucleon basis of the relativistic
Hartree-Bogoliubov (RHB) model. The
RHB model presents the relativistic extension of the Hartree-Fock-Bogoliubov
framework, and it provides a unified description of particle-hole 
($ph$) and particle-particle ($pp$) correlations. In this
framework the ground state of a nucleus can be written either in the
quasiparticle basis as a product of independent quasi-particle states, or in
the canonical basis as a highly correlated BCS-state. By definition, the
canonical basis diagonalizes the density matrix and it is always localized.
It describes both the bound states and the positive-energy single-particle
continuum. The formulation of the RQRPA in the canonical basis is
particularly convenient because, in order to describe transitions to
low-lying excited states in weakly bound nuclei, the
two-quasiparticle configuration space must include states with both nucleons
in the discrete bound levels, states with one nucleon in a bound level and
one nucleon in the continuum, and also states with both nucleons in the
continuum.

The relativistic QRPA of Ref.~\cite{Paa.03} is fully self-consistent.
For the interaction in the particle-hole channel effective Lagrangians with
nonlinear meson self-interactions are used, and pairing correlations are
described by the pairing part of the finite range Gogny interaction. Both in
the $ph$ and $pp$ channels, the same interactions are used in the RHB
equations that determine the canonical quasiparticle basis, and in the
matrix equations of the RQRPA. The RQRPA configuration space 
includes also the Dirac sea of negative energy states.
The RHB+RQRPA approach has been tested in the example of 
multipole excitations of neutron rich oxygen isotopes, and the model
has been employed in the analysis of the evolution of the low-lying 
isovector dipole strength in Sn isotopes and N=82 isotones. 

Relativistic mean-field and RPA calculations based on
effective Lagrangians with nonlinear meson self-interactions 
present not only a number of technical problems,
but also the description of finite nuclei obtained with these  
effective interactions is not satisfactory,            
especially for isovector properties. As an alternative, in a number of 
recent applications models with density-dependent meson-nucleon 
vertex functions have been used. 
Although the two classes of models are essentially based 
on the same microscopic structure, i.e. on density-dependent 
interactions, the latter can be more directly related to the
underlying microscopic nuclear interactions. Several   
recent analyses have shown that relativistic effective 
interactions with explicit density dependence of the meson-nucleon
couplings provide an improved description of asymmetric nuclear matter, 
neutron matter and nuclei far from stability. 
In Ref.~\cite{NVFR.02} we have extended the 
relativistic Hartree-Bogoliubov (RHB) model to include
density dependent meson-nucleon couplings. The effective Lagrangian is
characterized by a phenomenological density dependence of the $\sigma$,
$\omega$ and $\rho$ meson-nucleon vertex functions, adjusted to properties
of nuclear matter and finite nuclei. 
It has been shown that, in comparison 
with standard RMF effective interactions with nonlinear meson-exchange
terms, the new density-dependent meson-nucleon 
force DD-ME1 significantly
improves the description of asymmetric nuclear matter 
and of ground-state properties of $N\neq Z$ nuclei. This is,
of course, very important for the extension of RMF-based models 
to exotic nuclei far from $\beta$-stability, and
for applications in the field of nuclear astrophysics. 

In Ref.~\cite{NVR.02} the matrix equations of the 
relativistic random-phase approximation
(RRPA) have been derived for an effective Lagrangian characterized by 
density-dependent meson-nucleon couplings. The explicit
density dependence introduces 
rearrangement terms in the residual two-body interaction, and their 
contribution is essential for a quantitative description of excited
states. Illustrative calculations of the isoscalar monopole,
isovector dipole and isoscalar quadrupole response of $^{208}$Pb,
were performed in the fully self-consistent RRPA framework based
on effective interactions with a phenomenological density 
dependence adjusted to nuclear matter and ground-state properties
of spherical nuclei. 

In this work we extend the model developed in Ref.~\cite{Paa.03},
and formulate a proton-neutron relativistic QRPA in the canonical 
single-nucleon basis of the RHB model. The proton-neutron RQRPA, 
with the density-dependent effective interaction DD-ME1, will
be employed in the analysis of nuclear spin and isospin excitations. 

Collective spin and isospin excitations in atomic nuclei have 
been the subject of many experimental and theoretical studies
(for an extensive review see Ref.~\cite{Ost.92}).
Nucleons with spin up and spin down can oscillate either in 
phase (spin scalar S=0 mode) or out of phase (spin vector S=1 
mode). The spin vector, or spin-flip excitations can be of 
isoscalar (S=1, T=0) or isovector (S=1, T=1) nature. These 
collective modes provide direct information on the spin and
spin-isospin dependence of the effective nuclear interaction.

Especially interesting is the collective spin-isospin oscillation 
with the excess neutrons coherently changing the direction of their 
spins and isospins without changing their orbital motion -- the 
Gamow-Teller resonance (GTR) $J^\pi = 1^{+}$. This collective mode 
was predicted already in 1963 \cite{Ike.63}, but it was 
only in 1975 that the first experimental indications of the 
GT resonance were observed in $(p,n)$ charge-exchange reactions at
intermediate energies \cite{Doe.75}. The detailed knowledge of  
GT strength distributions is essential for 
the understanding of nuclear reactions in the process of nucleosynthesis. 
In particular, the low-lying GT strength is directly related to  
$\beta$-decay rates, as well as to the electron-capture 
process leading to the
stellar collapse and supernovae explosion. 
In many nuclei all over the periodic table the
GT strength distribution, when measured in the excitation energy region 
where the most pronounced GT peaks occur, is quenched by more than
20\% when compared to a model independent sum rule. Two physically 
different mechanisms had been suggested as a possible 
explanation of the quenching of the total GTR strength:
(i) nuclear configuration mixing -- the high-lying 
$2p-2h$ states mix with the $1p-1h$ GT states and 
shift the GT strength to high-energy region 
far beyond the resonance \cite{Shi.74,Ber.82,DOS.86};
(ii) the high-energy $\Delta$-isobar -- nucleon-hole 
configurations ($\Delta-h$) couple to the GT mode and 
remove part of the strength from the low-lying 
excitation spectrum \cite{Eric.73,Knu.80}. Recent $(p,n)$ scattering 
experiments have shown, however, 
that only a small fraction of the GT quenching originates
from $\Delta-h$ transitions \cite{Wak.97}. 

The GT spin-flip isovector excitations and the 
related $\beta$-decays have 
been theoretically investigated by employing: (i) the shell-model approach,
(ii) the non-relativistic proton-neutron 
quasiparticle random phase approximation, and (iii) the relativistic
random phase approximation in closed-shell nuclei.
Experimental data on charge-exchange excitations in light and medium-mass
nuclei are very successfully reproduced by large-scale shell-model
calculations. However, as the number of valence nucleons 
increases, the dimension of shell-model configuration space
becomes far too large for practical calculations. Shell-model 
calculations have been recently 
extended to the region of $pf$-shell nuclei with $A=45-65$.
Experimental Gamow-Teller strength distributions and
nuclear $\beta$-decay half-lives have been 
reproduced~\cite{Cau.99}. The GT response in medium-mass nuclei
has also been successfully described by shell-model Monte Carlo
calculations (SMMC) \cite{Rad.97}.

The proton - neutron quasiparticle random phase approximation
(PN-QRPA) can be used to describe nuclei in
mass regions that are presently beyond the reach of the most sophisticated
shell-model calculations. The pioneering work 
of Halbleib and Sorensen~\cite{Hal.67} was based on simple separable forces.
Further developments employed zero-range interactions with BCS-type 
pairing. Most studies of GT excitations have been based on
Skyrme interactions \cite{Gia.81,Col.94,Ham.93,Suz.00,Ben.02}.
It has also been shown that the inclusion of particle-particle 
correlations in the QRPA
residual interaction has an important effect in the calculation of 
$\beta$-decay rates~\cite{Cha.83,Che.95,Eng.88,Bor.95,Bor.96,Eng.99}. 

Experimental data on unstable nuclei close to N=Z line, that became 
available in recent studies with radioactive beams, have also 
renewed the interest in the role of proton-neutron pairing.  
In medium-heavy and heavy nuclei closer 
to the line of $\beta$-stability, the proton and neutron Fermi levels
are located in different major shells, and therefore the 
contribution of proton-neutron pairing correlations to the ground-state 
binding is usually neglected. The low-lying GT strength responsible 
for $\beta$-decay, however, involves 
proton-neutron particle-particle correlations
and is sensitive to the $T=0$ pairing interaction. In Ref.~\cite{Eng.99}  
$\beta$-decay rates for spherical neutron-rich $r$-process waiting-point 
nuclei have been calculated within a fully self-consistent QRPA, 
formulated in the Hartree-Fock-Bogoliubov canonical single-particle basis.
It has been shown that the proton-neutron particle-particle interaction 
has a pronounced effect on the calculated half-lives.

Surprisingly little work has been reported on the description 
of charge-exchange excitations in finite nuclei in the framework  
of relativistic mean-field models. The first
relativistic RPA calculations of isobaric analog resonances (IAR)
and Gamow-Teller
resonances have been performed only recently 
in Refs.~\cite{Con.98,Con.00}. 
This analysis was, however, 
restricted to doubly closed-shell nuclei. A rather small 
configuration space was used and, furthermore, configurations that 
include empty states from the negative-energy Dirac sea were 
neglected. In a very recent work \cite{Kur.03a} it has been shown, on the
other hand, that the inclusion of negative-energy Dirac sea states has a 
pronounced effect on the calculated Gamow-Teller sum rule.
A more complete relativistic RPA calculation of  
GT resonances in doubly closed-shell nuclei has been reported in 
Ref.~\cite{Ma.03}. The ground states of 
$^{48}$Ca, $^{90}$Zr and $^{208}$Pb were calculated in the 
relativistic mean-field (RMF) model with the NL3 effective interaction.
The spin-isospin correlations in the RRPA calculation of the GT 
response functions were induced by the isovector mesons $\pi$ and $\rho$.
In addition to the standard non-linear NL3 effective interaction \cite{LKR.97}
with the vector rho-nucleon coupling,
the effective Lagrangian included the pseudo-vector pion-nucleon 
interaction. Although in the relativistic mean-field
description of the nuclear ground state the direct one-pion 
contribution vanishes at the Hartree level because of parity conservation, 
the pion nevertheless plays an important role for excitations 
that involve spin degrees of freedom. Since it has a relatively small mass, 
the pion mediates the effective nuclear interaction over large distances. 
The $p-h$ residual interaction with $\rho$- and $\pi$-meson exchange has 
been extensively used in 
non-relativistic RPA calculations of charge-exchange 
excitations \cite{Kre.81,Kre.88}. 
Because of the derivative type of the pion-nucleon coupling, it is 
also necessary to include a zero-range Landau-Migdal term that
accounts for the contact part of the nucleon-nucleon interaction.  
The analysis of Ref.~\cite{Ma.03} has shown that 
the RRPA calculation with the NL3 
effective interaction, the pseudo-vector pion-nucleon coupling 
($m_{\pi}=138$ MeV and $f_{\pi}^{2}/{4\pi}=0.08$), and 
the Landau-Migdal force with the strength parameter $g_{0}^{\prime}= 0.6$,
reproduces the experimental excitation energies of 
the main components of the GT resonances in $^{48}$Ca, $^{90}$Zr 
and $^{208}$Pb.  

In Ref.~\cite{VPNR.03} we have calculated the GTR and IAR for a sequence
of even-even Sn target nuclei by using the 
framework of the relativistic
Hartree-Bogoliubov model plus proton-neutron quasiparticle random-phase
approximation. The calculation reproduces the experimental data on 
ground-state properties, as well as the excitation energies of 
the isovector excitations. It has been shown that the isotopic dependence of 
the energy spacings between the GTR and IAR provides direct information 
on the evolution of neutron skin-thickness along the Sn 
isotopic chain. A new method has been suggested for 
determining the difference 
between the radii of the neutron and proton density distributions along 
an isotopic chain, based on measurement of the excitation energies 
of the GTR relative to the IAR.

This work presents a much more detailed analysis of charge-exchange modes,
both for doubly closed-shell and for open-shell nuclei.
In Sec.~\ref{secII} we introduce the formalism. The matrix
equations of the proton-neutron relativistic QRPA are formulated in 
the canonical basis of the
relativistic Hartree-Bogoliubov (RHB) model for
spherical nuclei. In Sec.~\ref{secIII} the RHB+RQRPA model is employed
in illustrative calculations of charge-exchange collective modes:
isobaric analog resonances and Gamow-Teller resonances. 
Section~\ref{secIV} contains the conclusions.


\section{\label{secII}Proton-neutron relativistic quasiparticle
random phase approximation}

In Ref.~\cite{Paa.03} we have derived
the relativistic quasiparticle random phase approximation (RQRPA) 
from the time-dependent relativistic Hartree-Bogoliubov (RHB) model
in the limit of small amplitude oscillations.
The full RQRPA equations are rather complicated and
it is considerably simpler to solve these equations in the
canonical basis, in which the relativistic 
Hartree-Bogoliubov wave functions can be
expressed in the form of BCS-like wave functions. 
In this case one needs only the matrix
elements $V_{\kappa ^{{}}\lambda ^{\prime }\kappa ^{\prime }\lambda
^{{}}}^{ph}$ of the residual 
$ph$-interaction, and the matrix elements $V_{\kappa
^{{}}\kappa ^{\prime }\lambda ^{{}}\lambda ^{\prime }}^{pp}$ 
of the pairing $pp$-interaction, as well as certain combinations of the 
occupation factors $u_{\kappa }$, $v_{\kappa }$. The details of the RHB 
model, of the RQRPA formalism, and the technicalities of the 
solution of the RQRPA equations in the canonical basis, 
are given in Ref.~\cite{Paa.03}. 
In this section we only collect those expressions which are 
specific for the proton-neutron
relativistic quasiparticle random phase approximation (PN-RQRPA).
The model can be considered as a relativistic extension of 
the fully-consistent proton-neutron QRPA formulated in Ref.~\cite{Eng.99},
and applied in an analysis of $\beta$-decay rates of $r$-process nuclei.

We consider transitions between the $0^+$ ground state of a spherical
even-even parent nucleus and the $J^\pi$ excited state of the 
corresponding odd-odd daughter
nucleus. These transitions are induced by a charge-exchange operator $T^{JM}$.
Taking into account the rotational invariance of the nuclear system, the
quasiparticle pairs can be coupled to good angular momentum and the matrix
equations of the PN-RQRPA read

\begin{equation} \left( \begin{array} [c]{cc}
A^{J} & B^{J}\\
B^{^{\ast}J} & A^{^{\ast}J}
\end{array}
\right)  \left( \begin{array} [c]{c}
X^{\lambda J}\\
Y^{\lambda J}
\end{array}
\right)  =E_{\lambda}\left( \begin{array} [c]{cc}
1 & 0\\
0 & -1
\end{array}
\right)  \left( \begin{array} [c]{c}
X^{\lambda J}\\
Y^{\lambda J}
\end{array}\right) \; . 
\label{pnrqrpaeq}
\end{equation}
The matrices $A$ and $B$ are defined in the canonical basis \cite{Rin.80}
\begin{eqnarray}
A_{pn,p^\prime n^\prime}^{J} &=& H^{11}_{pp^\prime}\delta_{nn^\prime} +
  H^{11}_{nn^\prime}\delta_{pp^\prime}  \nonumber \\ & & +
\lp( u_p v_n u_{p^\prime} v_{n^\prime} + v_p u_n v_{p^\prime} u_{n^\prime}\rp)
 V_{pn^\prime n p^\prime}^{ph J} + 
\lp( u_p u_n u_{p^\prime} u_{n^\prime} + v_p v_n v_{p^\prime} v_{n^\prime}\rp) 
 V_{pn p^\prime n^\prime}^{pp J} \nonumber \\
B_{pn,p^\prime n^\prime}^{J} &=& (-1)^{j_{p^\prime}-j_{n^\prime}+J}
\lp( u_p v_n v_{p^\prime} u_{n^\prime} + v_p u_n u_{p^\prime} v_{n^\prime}\rp)
 V_{pp^\prime n n^\prime}^{ph J} \nonumber \\ & &- 
\lp( u_p u_n v_{p^\prime} v_{n^\prime} + v_p v_n u_{p^\prime} u_{n^\prime}\rp) 
 V_{pn p^\prime n^\prime}^{pp J} \; .
\label{abmat}
\end{eqnarray}
Here $p$, $p^\prime$, and $n$, $n^\prime$ denote proton and neutron quasiparticle
canonical states, respectively, 
$V^{ph}$ is the proton-neutron particle-hole residual
interaction, and $V^{pp}$ is the corresponding particle-particle interaction.
The canonical basis diagonalizes the density matrix and 
the occupation amplitudes $v_{p,n}$ are the corresponding eigenvalues. 
The canonical basis, however, does not diagonalize the Dirac single-nucleon
mean-field Hamiltonian $\hat{h}_{D}$ and the pairing field $\hat{\Delta}$,
and therefore the off-diagonal matrix elements $H^{11}_{nn^\prime}$ and
$H^{11}_{pp^\prime}$ appear in Eq. (\ref{abmat}):
\begin{equation}
H_{\kappa \kappa^\prime}^{11}=(u_{\kappa }u_{\kappa^\prime }
-v_{\kappa }v_{\kappa^\prime
})h_{\kappa \kappa^\prime }-(u_{\kappa }v_{\kappa^\prime }+
v_{\kappa }u_{\kappa^\prime
})\Delta _{\kappa \kappa^\prime }\;,
\label{H11}
\end{equation}
For each energy $E_{\lambda}$, $X^{\lambda J}$ and $Y^{\lambda J}$
in Eq. (\ref{abmat})
denote the corresponding forward- and backward-going QRPA amplitudes,
respectively. 
The total strength for the transition between the ground state of the even-even
(N,Z) nucleus and the excited state of the odd-odd (N+1,Z-1) or (N-1,Z+1) 
nucleus, induced by the operator $T^{JM}$, reads
\begin{equation}
B_{\lambda J}^{\pm} = \lp| \sum_{pn} <p||T^J||n> 
\lp( X_{pn}^{\lambda J} u_p v_n + (-1)^J Y_{pn}^{\lambda J}v_p u_n \rp) \rp|^2
\; .
\label{strength-}
\end{equation}
The discrete strength distribution is folded by the Lorentzian function
\begin{equation}
R(E)^{\pm} = \sum_{\lambda}B_{\lambda J}^{\pm}\frac{1}{\pi}
\frac{\Gamma/2}{(E-E_{\lambda_{\pm}})^2+(\Gamma /2)^2} \; .
\label{lorentzian}
\end{equation}
In the illustrative
calculations included in the following sections, the choice for
the width of the Lorentzian function is 1 MeV.

The spin-isospin dependent interaction terms are generated by the $\rho$-
and $\pi$-meson exchange. Because of parity conservation, 
the one-pion direct contribution vanishes 
in the mean-field calculation of a nuclear ground state. Its inclusion is 
important, however, in calculations of excitations that involve
spin and isospin degrees of freedom.
The particle-hole residual interaction in the PN-RQRPA is derived from the 
Lagrangian density 
\begin{equation}
\mathcal{L}_{\pi + \rho}^{int} = 
      - g_\rho \bar{\psi}\gamma^{\mu}\vec{\rho}_\mu \vec{\tau} \psi 
      - \frac{f_\pi}{m_\pi}\bar{\psi}\gamma_5\gamma^{\mu}\partial_{\mu}
        \vec{\pi}\vec{\tau} \psi \; . 
\label{lagrres}	
\end{equation}
Vectors in isospin space are denoted by arrows, and boldface symbols 
will indicate vectors in ordinary three-dimensional space.

The coupling between the $\rho$-meson and the nucleon is assumed to be a vertex
function of the vector density $\rho_v = \sqrt{j_\mu j^\mu}$, with 
$j_\mu = \bar{\psi}\gamma_\mu\bar{\psi}$. In Ref.~\cite{NVR.02} it has been
shown that the explicit density dependence of the meson-nucleon couplings
introduces additional 
rearrangement terms in the residual two-body interaction of the RRPA, 
and that their 
contribution is essential for a quantitative description of excited
states. However, since the rearrangement terms include the corresponding 
isoscalar ground-state densities, it is easy to see that they are absent in the 
charge exchange channel, and the residual two-body interaction reads
\begin{eqnarray}
V(\bm{r}_1,\bm{r}_2) &=& \vec{\tau}_1\vec{\tau}_2 (\beta \gamma^\mu)_1
      (\beta \gamma_\mu)_2 g_\rho(\rho_v(\bm{r}_1)) g_\rho(\rho_v(\bm{r}_2))
      D_\rho (\bm{r}_1,\bm{r}_2) \nonumber \\
      &-&\left(\frac{f_\pi}{m_\pi}\right)^2\vec{\tau}_1\vec{\tau}_2
      (\bm{\Sigma}_1\bm{\nabla}_1)(\bm{\Sigma}_2\bm{\nabla}_2)
      D_\pi (\bm{r}_1,\bm{r}_2)\;.
\end{eqnarray}
$D_{\rho (\pi)}$ denotes the meson propagator
\begin{equation}
 D_{\rho (\pi)} = \frac{1}{4\pi}
        \frac{e^{-m_{\rho (\pi)}|\bm{r}_1-\bm{r}_2|}}{|\bm{r}_1-\bm{r}_2|}\;, 
\end{equation}	   

and 
\begin{equation}
\bm{\Sigma} = \left(
\begin{array}
[c]{cc}%
\bm{\sigma} & 0\\
0 & \bm{\sigma}%
\end{array}
\right)  \;. \label{bigsigma}%
\end{equation}

For the $\rho$-meson coupling we adopt the functional form used in
the DD-ME1 density-dependent effective interaction~\cite{NVFR.02}
\begin{equation}
g_\rho (\rho_v) = g_\rho(\rho_{sat})exp[-a_\rho (x-1)]\;,
\end{equation}
where $x=\rho_v / \rho_{sat}$, and $\rho_{sat}$ denotes the saturation vector 
nucleon density in symmetric nuclear matter.
For the pseudovector pion-nucleon coupling we use the standard values 
\begin{equation}
m_{\pi}=138.0~MeV~~~~\;\;\;\;\frac{\;f_{\pi}^{2}}{4\pi}=0.08\;.
\end{equation}
The derivative type of the pion-nucleon coupling necessitates the  
inclusion of the zero-range
Landau-Migdal term, which accounts for the contact part of the 
nucleon-nucleon interaction
\begin{equation}
V_{\delta\pi} = g^\prime \lp( \frac{f_\pi}{m_\pi} \rp)^2 
\vec{\tau}_1\vec{\tau}_2 \bm{\Sigma}_1 \cdot \bm{\Sigma}_2 
\delta (\bm{r}_1-\bm{r}_2)\; ,
\label{deltapi}
\end{equation} 
with the parameter 
$g^{\prime} \approx 0.6$ adjusted
to reproduce experimental data on the GTR excitation energies. 

With respect to the RHB calculation of the ground state of an even-even nucleus, 
the charge-exchange channel includes the additional one-pion exchange 
contribution. The PN-RQRPA model is fully consistent:
the same interactions, both in the particle-hole and particle-particle
channels, are used in the RHB equation that determines the
canonical quasiparticle basis, and in the PN-RQRPA equation (\ref{pnrqrpaeq}). 
In both channels the same strength parameters of the interactions are used in
the RHB and RQRPA calculations.  

The two-quasiparticle configuration space includes states with
both nucleons in the discrete bound levels, states with one nucleon in the
bound levels and one nucleon in the continuum, and also states with both
nucleons in the continuum. In addition to the configurations built from
two-quasiparticle states of positive energy, the RQRPA configuration space
contains pair-configurations formed from the fully or partially occupied
states of positive energy and the empty negative-energy states from the
Dirac sea. As will be shown in the next section, 
the inclusion of configurations built from occupied
positive-energy states and empty negative-energy states is essential for the
consistency of the model.

Nuclear properties calculated with the RHB plus RQRPA model 
depend on the choice of the effective RMF Lagrangian in the $ph$-channel,
as well as on the treatment of pairing correlations. In this work, for
the RHB calculation of ground states and in the $ph$-channel of the residual 
interaction, we use the density-dependent effective interaction DD-ME1. 
In Ref.~\cite{NVFR.02} the RHB model
with the density-dependent interaction DD-ME1 in the
$ph$-channel, and with the finite range Gogny interaction D1S in
the $pp$-channel, has been tested
in the analysis of the equations of state for symmetric
and asymmetric nuclear matter, and of ground-state properties of the
Sn and Pb isotopic chains. It has been shown that, as compared to
standard non-linear relativistic mean-field effective forces,
the interaction DD-ME1 has better isovector properties and therefore
provides an improved description of asymmetric nuclear matter, neutron
matter and nuclei far from stability.

In the $pp$-channel of the RHB model we have used a
phenomenological pairing interaction, the pairing part of the Gogny force,
\begin{equation}
V^{pp}(1,2)~=~\sum_{i=1,2}e^{-((\mathbf{r}_{1}-\mathbf{r}_{2})/{\mu _{i}}%
)^{2}}\,(W_{i}~+~B_{i}P^{\sigma }-H_{i}P^{\tau }-M_{i}P^{\sigma }P^{\tau }),
\label{Gogny}
\end{equation}
with the set D1S \cite{Ber.84} for the parameters $\mu _{i}$, $W_{i}$, $%
B_{i} $, $H_{i}$ and $M_{i}$ $(i=1,2)$. This force has been very carefully
adjusted to the pairing properties of finite nuclei all over the periodic
table. In particular, the basic advantage of the Gogny force is the finite
range, which automatically guarantees a proper cut-off in momentum space.
In the present analysis we will also use the Gogny interaction
in the $T=1$ $pp$-channel of the PN-RQRPA. 
For the $T=0$ proton-neutron pairing interaction in open shell nuclei 
we employ a similar interaction:
a short-range repulsive Gaussian combined with a
weaker longer-range attractive Gaussian:
\begin{equation}
\label{eq2}
V_{12}
= - V_0 \sum_{j=1}^2 g_j \; {\rm e}^{-r_{12}^2/\mu_j^2} \;
    \hat\Pi_{S=1,T=0}
\quad ,
\label{pn-pair}
\end{equation}
where $\hat\Pi_{S=1,T=0}$ projects onto states with $S=1$ and $T=0$.  
This interaction was used in the non-relativistic QRPA calculation
\cite{Eng.99} of $\beta$-decay rates for spherical neutron-rich 
$r$-process waiting-point
nuclei. As it was done in Ref.~\cite{Eng.99}, we take
the ranges $\mu_1$=1.2\,fm and $\mu_2$=0.7\,fm of the two Gaussians from the
Gogny interaction (\ref{Gogny}), and choose the relative strengths $g_1 =1$ and
$g_2 = -2$ so that the force is repulsive at small distances.  The only
remaining free parameter is $V_0$, the overall strength.


\section{\label{secIII} Charge-exchange collective modes of excitations}

\subsection{The isobaric analog resonance}

The isobaric analog resonance (IAR) presents the simplest 
charge-exchange excitation mode, detected already forty years 
ago in experiments on low-energy proton elastic scattering from 
heavy nuclei. The observed resonances were at energies consistent
with the interpretation that they were isobaric analog states in 
the compound nucleus. One of the important characteristics of an
IAR is its narrow width. This is because it has the same isospin as
the parent state, while the neighboring states have the isospin of 
the ground state of the daughter nucleus, i.e. they differ in 
isospin by one unit. This means that they will couple only weakly
with the IAR.

As a first test of our PN-RQRPA model, we calculate the IAR strength
functions, i.e. the distribution of $J^{\pi}=0^{+}$. The Fermi 
transition operator reads
\begin{equation}
T_{\beta^{\pm}}^{F}=\sum_{i=1}^{A}\tau_{\pm}.
\label{iaroperator}
\end{equation}
In Fig. \ref{figA} we display the PN-RRPA response to the operator
(\ref{iaroperator}) for $^{48}$Ca, $^{90}$Zr and $^{208}$Pb. 
The strength distributions are dominated by a single
IAR peak, which corresponds to a coherent superposition of $\pi p-\nu h$ (or
proton-neutron 2qp) excitations. The calculated excitation energies 
(evaluated with respect to the ground state of the parent nucleus) are 
compared with the corresponding experimental values (thick arrows) from 
$(p,n)$ scattering data for
$^{48}$Ca~\cite{And.85}, $^{90}$Zr~\cite{Bai.80,Wak.97}, and $^{208}%
$Pb~\cite{Aki.95}. The agreement between the PN-RRPA and experimental 
data is indeed very good. 
We have also verified that the calculated strength distributions exhaust
the Fermi non-energy weighted sum rule 
\begin{equation}
S_{\beta^{-}}(F)=N-Z. \label{ikedaiar}%
\end{equation}

The proton-neutron relativistic QRPA provides a natural framework
for the description of spin and isospin excitations in open-shell nuclei.
In Fig.~\ref{figB} we plot the calculated IAR excitation energies for the
sequence of even-even Sn target nuclei with $A=108 - 132$. The 
result of fully self-consistent RHB plus proton-neutron RQRPA 
calculations are shown in comparison with experimental data 
obtained in a systematic study of the ($^3$He,t) charge-exchange 
reaction over the entire range of stable Sn isotopes \cite{Pham.95}.
The calculated values reproduce the empirical 
mass dependence of the IAR excitation energies, and 
we also notice a very good agreement with available experimental data. 
For the nuclei with
$A=112 - 124$ the largest difference between the theoretical 
and experimental IAR excitation energies is $\approx 200$ keV. 
For the Sn nuclei with $A=108 - 132$, in Figs.~\ref{figC} and \ref{figD}
we display the corresponding theoretical strength distribution functions.
To illustrate the importance of a consistent treatment 
of pairing correlations, in addition to the 
response functions calculated with the fully self-consistent RHB 
plus proton-neutron RQRPA, we also display the strength functions
obtained without including the proton-neutron residual pairing interaction. 
Proton-neutron pairing, of course, does not contribute in the 
RHB calculation of the ground states.
The dashed curves are calculated by including only the $ph$-channel
of the RQRPA residual interaction. In both cases 
(solid and dashed curves) $T=1$ neutron-neutron pairing has been 
included in the RHB calculations of the ground-states of Sn target
nuclei. The results shown in Figs.~\ref{figC} and \ref{figD} clearly
illustrate the pronounced effect that $T=1$
proton-neutron pairing correlations have on the calculated IAR.
Without proton-neutron pairing the calculated IAR excitation energies
for nuclei with $A \geq 116$ are simply too high.
The attractive proton-neutron pairing interaction 
redistributes the strength between
various QRPA components, and lowers the calculated 
excitation energy to the value which corresponds to the 
difference between the Coulomb energies of the target nucleus
and the $(Z+1,N-1)$ daughter nucleus.  
As one would expect, the effect gradually weakens
as the closed shell at $N=82$ is approached.  
The effect of the proton-neutron residual pairing  
on the IAR strength distributions in 
lighter Sn isotopes with $A\leq 114$ is more dramatic. Without the   
contribution of the $pp$-channel to the proton-neutron residual interaction, 
the response function is fragmented 
and it does not display the experimentally observed single narrow 
resonance. Only with the inclusion of proton-neutron pairing a 
single collective state is obtained. 

The mechanism through which the PN-RQRPA builds the IAR in Sn isotopes 
is illustrated in Fig.~\ref{figE} for 
$^{108}$Sn, $^{114}$Sn and $^{120}$Sn. The figure displays the 
discrete QRPA spectra for the following three cases. 
In the panels on the left 
we plot the spectra calculated without any pairing interaction, 
either in calculation of the ground-state, or in the 
residual proton-neutron interaction. It is immediately clear why
the three particular examples were selected. Without pairing, 
the $g_{7/2}$ neutron orbital is fully occupied in the ground-state
of $^{108}$Sn, the $g_{7/2}$ and the $d_{5/2}$ orbitals are 
fully occupied in the ground-state of $^{114}$Sn, and 
for $^{120}$Sn all the neutron orbitals in the shell $N=50-82$ are 
occupied, except $h_{11/2}$. For all three nuclei the response function
to the Fermi operator (\ref{iaroperator}) displays a single peak at 
the excitation energy very close to the experimental position of the 
IAR. The distribution of neutron-to-proton
RPA amplitudes is shown next to each major peak. For example, in the 
case of $^{108}$Sn 99\% of the strength results from the transition 
$\nu~g_{7/2}$ $\rightarrow$ $\pi~g_{7/2}$, whereas for $^{120}$Sn
98\% of the strength is distributed between the four orbitals 
occupied by the neutrons in the $N=50-82$ shell.
In the next step (middle panels in Fig.~\ref{figE}) we include the Gogny
neutron-neutron pairing in the RHB calculation of the ground states, 
and as a result all the $N=50-82$ neutron 
orbitals acquire finite occupation probabilities.
The $pp$-channel, however, is not allowed to contribute to the 
QRPA matrix elements of the proton-neutron residual interaction. 
In this case the QRPA spectra become fragmented in $^{108}$Sn and  
$^{114}$Sn. Considerable strength is transferred to states
which correspond to transitions from neutron orbitals with 
low occupation probabilities: $s_{1/2}$ and $h_{11/2}$. 
For the most pronounced peaks we have included
the distribution of neutron-to-proton QRPA amplitudes.The 
fragmentation does not occur in $^{120}$Sn where, as a result 
of ground-state pairing correlations, all the $N=50-82$ neutron 
orbitals have considerable occupation probabilities. In 
$^{120}$Sn we notice only a redistribution of QRPA amplitudes 
in favor of higher-lying orbitals,
and consequently the repulsive particle-hole residual interaction
shifts the IAR to higher energy.
For the two nuclei from the lower half of the $N=50-82$ neutron shell,
the $ph$ residual interaction
fragments the QRPA spectra and effectively precludes the formation
of the IAR. The panels on the right-hand side of 
Fig.~\ref{figE} display the result
obtained by fully-consistent PN-RQRPA calculations, including the 
$T=1$ proton-neutron pairing channel: the strength has been collected in 
a single peak -- the IAR. We notice that with
respect to the case (a) without any pairing correlations,
the IAR strength has
been redistributed among the neutron-to-proton QRPA amplitudes,
reflecting the occupation of higher-lying neutron orbitals
in the parent nucleus. The calculated
excitation energies, as we have shown in  Fig.~\ref{figB}, are in very good
agreement with the experimental data on the IAR in Sn nuclei \cite{Pham.95}. 
The total effect is, of course, simply a consequence of the fact that 
the pairing interaction is isospin invariant, and therefore commutes
with the Fermi operator (\ref{iaroperator}). However,
the results shown in Figs.~\ref{figC} - \ref{figE} illustrate the 
importance of a consistent treatment of pairing correlations by including
an isospin invariant pairing interaction in the 
description of the ground state and the dynamics of IAR in $N\neq Z$ nuclei.
\subsection{The Gamow-Teller resonance}

%
The Gamow-Teller resonance represents a coherent superposition of 
high-lying $J^{\pi}=1^{+}$ proton-particle -- neutron-hole configurations 
of maximum collectivity associated with charge-exchange excitations 
of neutrons from orbitals with $j = l + \frac{1}{2}$ into proton 
orbitals with $j = l - \frac{1}{2}$. The GT operator reads
\begin{equation}
T_{\beta^{\pm}}^{GT}=\sum_{i=1}^{A}\bm{\Sigma}\tau_{\pm} \; .
\label{gtopera}
\end{equation}
The calculated GT strength distributions for 
$^{48}$Ca, $^{90}$Zr and $^{208}$Pb
are shown in Fig.~\ref{figF}. In addition to the high-energy GT 
resonance -- a collective superposition of direct spin-flip 
($j = l + \frac{1}{2}$ $\rightarrow$ $j = l - \frac{1}{2}$) 
transitions, the response functions display a concentration of
strength in the low-energy tail. The transitions in the low-energy
region correspond to core-polarization 
($j = l \pm \frac{1}{2}$ $\rightarrow$ $j = l \pm \frac{1}{2}$), 
and back spin-flip 
($j = l - \frac{1}{2}$ $\rightarrow$ $j = l + \frac{1}{2}$)
neutron-hole -- proton-particle excitations.

As in the calculation of the IAR, the effective RMF interaction is
DD-ME1. In addition, the standard values have been used for the 
parameters of the pion-nucleon interaction Lagrangian: 
$m_{\pi}=138$ MeV and $f_{\pi}^{2}/{4\pi}=0.08$, and the parameter 
of the zero-range Landau-Migdal force $g^{\prime}= 0.55$ has 
been adjusted to reproduce the excitation energy of the GT
resonance in $^{208}$Pb. The calculated resonances are shown 
in comparison with experimental data (thick arrows) for the 
GTR excitation energies in $^{48}$Ca~\cite{And.85}, 
$^{90}$Zr~\cite{Bai.80,Wak.97}, and $^{208}$Pb~\cite{Hor.80,Aki.95, Kra.01}.
Although the residual interaction has been adjusted to reproduce
the GTR excitation energy in $^{208}$Pb, we notice a very good
agreement with experiment also for $^{48}$Ca and $^{90}$Zr. 
The adjusted value of $g^{\prime}$ in general depends on the choice
of the RMF effective interaction. In the RRPA calculation 
with the NL1~\cite{RRMGF.86} effective interaction of Ref.~\cite{Con.98}, 
the GTR excitation energies of $^{48}$Ca, $^{90}$Zr 
and $^{208}$Pb were best reproduced by $g^{\prime}= 0.7$, whereas
approximately the same quality of agreement with experimental data 
was obtained in Ref.~\cite{Ma.03} by using the NL3~\cite{LKR.97}  
effective interaction and $g^{\prime}= 0.6$. By using a set of
density-dependent RMF effective interactions from Ref.~\cite{NVR.02}
we have verified that, in general, there is a correlation between 
the value of the nuclear asymmetry energy at saturation $a_4$, and  
the value of $g^{\prime}$ adjusted to reproduce 
the GTR. In order to reproduce the experimental excitation energy
of the GTR, effective interactions with higher values of $a_4$ generally
require higher values of $g^{\prime}$.  

Gamow-Teller states calculated 
in relativistic models have recently attracted
considerable attention~\cite{Kur.03a,Kur.03b,Ma.03,Kur.04}. In 
particular, it has been shown that in a relativistic RPA calculation
the total GT strength in the nucleon sector is reduced by 
$\approx 12$\% in nuclear matter, and by $\approx 6$\% in finite 
nuclei when compared to the Ikeda sum rule \cite{Ike.63}
\begin{equation}
\left(  S_{\beta^{-}}^{GT}-S_{\beta^{+}}^{GT}\right)  =3(N-Z),\label{gtsrule}%
\end{equation}
where $S_{\beta^{\pm}}^{GT}$ denotes the total sum of Gamow-Teller
strength for the $\beta^{\pm}$ transition. The reduction has been attributed
to the effect of Dirac sea negative-energy states, i.e. the missing 
part of the sum rule is taken by particle-hole excitations formed from 
ground-state configurations of occupied 
states in the Fermi sea and empty negative-energy states in the Dirac sea.
The effect is illustrated in Fig.~\ref{figG}, where we display the 
running sum of GTR strength for $^{208}$Pb. The horizontal 
dotted line corresponds to the value $3(N-Z) = 132$ of the Ikeda sum rule.
The solid and dashed lines corresponds the values of the GTR sum 
calculated from $- \infty$ to the excitation energy denoted on the
abscissa. The big jump in the calculated GTR sum occurs, of course,
when the main GTR peak at $\approx 19$ MeV is included. 
The RRPA calculation represented by the dashed line includes
only positive energy $ph$ configurations. Even extending the sum 
up to 60 MeV, the total sum amounts only 
to $\approx 122$, that is 8\% less
than the Ikeda sum rule. The total sum rule $3(N-Z)$ is exhausted
by the calculated GT strength only when the relativistic RPA/QRPA 
space contains $ph$ excitations formed from ground-state 
configurations of the fully or partially occupied states of
positive energy, and the empty negative-energy states from the Dirac sea
(solid line in Fig.~\ref{figG}). The corresponding discrete GT spectrum is
shown in Fig.~\ref{figH}. Two regions of excitation energies are shown. 
The panel on the right-hand side contains the positive energy
$\pi p-\nu h$ strength, 
with a pronounced Gamow-Teller peak at $\approx 19$ MeV.
The left panel displays the negative energy spectrum 
built from $\pi\alpha-\nu h$ transitions ($\alpha$ denotes a 
negative energy state). Even though these transitions 
are much weaker than the GTR, there are many of them and 
their overall sum represents the strength missing  
in the positive energy sector.
Similar results for the GTR sum rule are also obtained for 
$^{48}$Ca and $^{90}$Zr.

For the case of open-shell nuclei we will compare the PN-RQRPA results
with experimental data on Gamow-Teller resonances obtained from 
Sn($^3$He,t)Sb charge-exchange reactions~\cite{Pham.95}. The GT 
strength distribution in $^{118}$Sn is shown in Fig.~\ref{figI}. 
Direct spin-flip transitions 
($\nu j = l + \frac{1}{2}$ $\rightarrow$ $\pi j = l - \frac{1}{2}$)
dominate the high-energy region above 10 MeV. The low-energy tail
of the strength distribution corresponds to core-polarization 
($\nu j = l \pm \frac{1}{2}$ $\rightarrow$ $\pi j = l \pm \frac{1}{2}$), 
and back spin-flip 
($\nu j = l - \frac{1}{2}$ $\rightarrow$ $\pi j = l + \frac{1}{2}$)
transitions. 

The solid curve in Fig.~\ref{figI} has been calculated without including 
the $T=0$ proton-neutron pairing in the RQRPA residual interaction.
The resulting high-energy GT strength displays pronounced fragmentation.
This fragmentation, however, is not induced by coupling to $2p-2h$ 
configurations (not included in the present version of the RQRPA), 
rather it is caused by the splitting between different $ph$ configurations.
GTR configuration splitting (an appearance of two or more collective
bumps with comparable intensities in the GTR strength function) was 
investigated in Ref.~\cite{GNU.89} in the framework of the shell 
optical model. For Sn nuclei, in particular, this effect was predicted
to occur as the valence neutron start to occupy the level with the 
highest $j$ in the shell -- $h_{11/2}$. The configuration splitting 
was attributed to the fact that the unperturbed energies
of the $(1g^\pi_{7/2})(1g^\nu_{9/2})^{-1}$ and 
$(1h^\pi_{9/2})(1h^\nu_{11/2})^{-1}$ configurations are almost degenerate.
The residual interaction removes this degeneracy and, as a result,
the main GT component separates in two distinct peaks. 
The ground-state pairing correlations have a strong influence
on the occupation of the $1h^\nu_{11/2}$ level, and therefore
the energy spacing between the two peaks will depend on 
$T=1$ pairing. For $^{118}$Sn
the calculated splitting of the GTR was 2.6 MeV~\cite{GNU.89}. 
Subsequently, the
fragmentation and splitting of the GTR in Sn nuclei was experimentally
investigated in Ref.~\cite{Pham.95}. The theoretically predicted 
configuration splitting of the main GT component, however, could not 
be observed, since the total widths of the resonances of $5-6$ MeV 
exceed the predicted splitting. The splitting of the main GT component 
shown in Fig.~\ref{figI} (solid line) of $\approx 3$ MeV is in 
agreement with the result of Ref.~\cite{GNU.89}. In addition, we 
find a third direct spin-flip component in the region above 10 MeV, 
predominately based on the configuration $(2d^\pi_{3/2})(2d^\nu_{5/2})^{-1}$.

The other curves shown in Fig.~\ref{figI} 
(dotted, dashed and dot-dashed) have 
been calculated with the inclusion of the $T=0$ proton-neutron pairing 
(\ref{pn-pair}) in the RQRPA residual interaction.
In Ref.~\cite{Eng.99} the overall strength parameter $V_0$ of the 
interaction was adjusted to measured half-lives of neutron-rich 
nuclei in regions where $r$-process path comes closest to the valley
of stability: $V_0 = 230$ MeV near $N=50$ and $V_0$ = 170 MeV near $N=82$. 
In the present analysis, however, $\beta$-decay 
half-lives are not calculated and we could not adjust 
$V_0$ in this way. The GT strength functions shown in Fig.~\ref{figI}
have been calculated for $V_0 =$ 200, 250 and 300 MeV, respectively.
As one would expect, the inclusion of $T=0$ pairing has a strong influence 
on the low-lying tail of the GT distribution, i.e. on the 
region that contributes to $\beta$-decay. Here, however, we are 
concerned with the main GT component. An important effect that we notice
is the disappearance of configuration splitting between the two high-energy
peaks. This happens because the $T=0$ pairing interaction does 
not affect configurations based on the $(1g^\nu_{9/2})$ orbital 
(fully occupied), whereas it lowers configurations based on 
$(1h^\nu_{11/2})$ and $(2d^\nu_{5/2})$ (partially occupied). Our calculation
therefore shows that the $T=0$ proton-neutron pairing strongly reduces the 
predicted configuration splitting of the main high-energy GT component.
In addition to the main GTR at $\approx 16$ MeV, however, part of the 
strength associated with direct spin-flip transitions is concentrated 
at $\approx 10$ MeV. Another interesting result is that the centroid of
the GT strength composed of direct spin-flip transitions 
($\nu j = l + \frac{1}{2}$ $\rightarrow$ $\pi j = l - \frac{1}{2}$) 
practically does not depend on the strength of the 
$T=0$ proton-neutron $pp$ residual interaction. This is shown in the 
insert of Fig.~\ref{figI}, where we plot the calculated centroid of 
the direct spin-flip GT strength, as function of the strength parameter 
$V_0$. Since we could not use experimental data on the GTR centroid
to adjust the strength of the $T=0$ pairing channel, the following 
illustrative calculations have been performed with $V_0 =$ 250 MeV.                                                                              
                                                                                
In a recent PN-RQRPA calculation~\cite{VPNR.03} we have  
shown that the isotopic dependence of 
the energy spacings between the GTR and IAR provides direct information 
on the evolution of neutron skin-thickness along the Sn 
isotopic chain. It has been suggested that the difference 
between the radii of the neutron and proton density distributions along 
an isotopic chain, could be determined 
from the measurement of the excitation energies 
of the GTR relative to the IAR. The calculation of Ref.~\cite{VPNR.03},
however, was performed with the NL3 relativistic mean-field 
effective interaction. Although NL3 has become a standard in RMF calculations, 
like many other non-linear meson-exchange effective interactions it 
has a high nuclear matter asymmetry energy, and consequently it
predicts rather large values for neutron density 
distribution radii. Here we repeat the 
calculation using the DD-ME1 effective interaction. Among other
ground-state nuclear properties, the parameters of DD-ME1 have also
been adjusted to experimental data on differences between radii of 
neutron and proton distributions. 
 
In Fig.~\ref{figJ} we display the calculated differences between 
the centroids of the direct spin-flip GT strength and
the respective isobaric analog resonances for the
sequence of even-even Sn target nuclei. For $A=112 - 124$
the results of RHB plus PN-RQRPA
calculation  (DD-ME1 density-dependent effective interaction,
Gogny $T=1$ pairing, $T=0$ pairing interaction (\ref{pn-pair}) 
with $V_0 =$ 250 MeV, the Landau-Migdal parameter $g^{\prime}= 0.55$), 
are compared with experimental data \cite{Pham.95}.
The calculated energy spacings are in very good agreement with 
the experimental values. As it has been emphasized in 
Ref.~\cite{VPNR.03}, the energy difference between the 
GTR and the IAS reflects the magnitude of the effective spin-orbit
potential. Many relativistic mean-field calculations have shown
that the magnitude of the 
spin-orbit potential is considerably reduced in neutron-rich  
nuclei \cite{LVR.97}, and a corresponding increase of the 
neutron skin has been predicted. 
In Fig.~\ref{figK} the calculated and experimental 
energy spacings between the 
GTR and IAS are plotted as a function of the calculated 
differences between the rms radii of the neutron and proton 
density distributions of even-even Sn isotopes (upper panel). 
Notice the uniform dependence of the energy 
difference between the GTR and IAS on the size of the neutron-skin.
This means that, in principle, the value of $r_n - r_p$ can be 
directly determined from the theoretical curve for a given value of 
$E_{\rm GT} - E_{\rm IAS}$.                                                                               
In the lower panel the calculated differences between neutron and  
proton rms radii are compared with available experimental data 
\cite{Kra.99}. The agreement between theoretical and                                                                                 
experimental values suggests that the
neutron-skin thickness can be determined from the measurement 
of the excitation energies of the GTR relative to IAS.                                                                                                                                                                                                                                                                                                                                                                                                       


\section{\label{secIV}Concluding remarks and outlook}

In this second part of the work on the relativistic 
quasiparticle random phase approximation (RQRPA),
we have formulated the 
proton-neutron RQRPA in the canonical single-nucleon 
basis of the relativistic Hartree-Bogoliubov (RHB) model, 
for an effective Lagrangian characterized by 
density-dependent meson-nucleon couplings.
The basic advantage of working in the canonical basis is
that, since it diagonalizes the density matrix, the 
canonical basis is always localized and so it can be used both
for bound states and for the positive energy single-particle 
continuum. This is especially important in the description of
low-lying excited states and giant resonances in weakly bound
nuclei far from stability. From the practical point of view, 
it is also considerably simpler to solve the RQRPA equations in the
canonical basis, in which the relativistic 
Hartree-Bogoliubov wave functions can be
expressed in the form of BCS-like wave functions. 

The inclusion of relativistic effective 
interactions with explicit density dependence of the meson-nucleon
couplings provides an improved description of asymmetric nuclear matter, 
nuclear ground-states and properties of excited states, 
especially in nuclei far from stability. 
The PN-RQRPA model includes both 
the $T=1$ and $T=0$ pairing channels, and the model space contains
pair configurations formed from the fully or 
partially occupied states of
positive energy in the Fermi sea, 
and the empty negative-energy states from the Dirac sea. 
It has been shown that 
the inclusion of configurations built from occupied
positive-energy states and empty negative-energy states 
is essential in order to satisfy model independent sum rules.

The proton-neutron relativistic QRPA provides a natural framework
for the description of spin and isospin excitations. In this work
the model has been  applied to the analysis of charge-exchange modes:
isobaric analog resonances and  Gamow-Teller resonances.
A very good agreement with experimental data has been obtained,
both for doubly-closed shell nuclei and for open shell nuclei. 
The importance of a consistent treatment of pairing correlations
has been demonstrated in the example of IAR and GTR in Sn isotopes.

In addition to the IA and GT $0\hbar \omega$ excitations, the (p,n) spectra 
show clear evidence for the collective spin-flip dipole 
($L=1, S=1$) $1\hbar \omega$ excitation and spin-flip quadrupole
($L=2, S=1$) $2\hbar \omega$ excitation. The broad ($ > 10$ MeV) spin-flip 
dipole and quadrupole resonances can be interpreted as superpositions 
of three collective modes with spin-parity $J^\pi = 0^-$, $1^-$, $2^-$, 
and $J^\pi = 1^+$, $2^+$, $3^+$, respectively. It has not yet been possible
to experimentally resolve these resonances into the different angular 
momentum components. Recently it has been suggested that experimental data
on charge-exchange spin-dipole resonances can be used to extract 
information about neutron-skin thickness in neutron-rich nuclei.

The PN-RQRPA will be applied to the study of spin-multipole charge-exchange 
modes. An interesting question is whether such a self-consistent approach,
which starts from the description of the nuclear ground-state, can 
simultaneously reproduce the dynamics of these high-lying and, 
in many cases, only weakly collective states. Another important topic 
will be the calculation of $\beta$-decay rates in semi-magic
neutron-rich nuclei. These nuclei are particularly important 
for $r$-process nucleosynthesys. Their lifetimes are rather 
long and they determine the abundances of stable nuclides. 
Very neutron-rich nuclei are also very difficult to reach
experimentally, and at present the only information on 
$\beta$-decay rates is provided by theoretical calculations.
This also emphasizes the importance of the comparison of the present approach  
with results of non-relativistic QRPA calculations.


\bigskip \bigskip

\leftline{\bf ACKNOWLEDGMENTS}

This work has been supported in part by the Bundesministerium
f\"ur Bildung und Forschung under project 06 TM 193, and by the
Gesellschaft f\" ur Schwerionenforschung (GSI) Darmstadt.
N.P. acknowledges support from the Deutsche
Forschungsgemeinschaft (DFG) under contract SFB 634.
==========================================================================

\newpage

\begin{figure}

\caption{PN-RQRPA $J^\pi=0^+$ strength distributions. 
The excitations of the isobaric analog resonances 
are compared with  experimental data (thick arrows) 
for $^{48}$Ca~\protect\cite{And.85}, $^{90}$Zr~\protect\cite{Bai.80,Wak.97}, 
and $^{208}$Pb~\protect\cite{Aki.95}.}
\label{figA}
\end{figure}

\begin{figure}
\caption{ RHB plus proton-neutron RQRPA results for the 
isobaric analog resonances of the 
sequence of even-even Sn target nuclei. 
The experimental data are from Ref.~\protect\cite{Pham.95}.}
\label{figB}
\end{figure}

\begin{figure}
\caption{PN-RQRPA $J^\pi=0^+$ strength distributions 
for even-even Sn target nuclei with $A= 108-118$. 
For description see text.}
\label{figC}
\end{figure}

\begin{figure} 
\caption{Same as in Fig.~\protect\ref{figC}, 
for even-even Sn target nuclei with $A= 120-132$.}
\label{figD}
\end{figure}

\begin{figure} 
\caption{PN-RQRPA $J^\pi=0^+$ discrete spectra for
$^{108}$Sn, $^{114}$Sn and $^{120}$Sn. The panels on the left 
correspond to calculations performed without including pairing correlations.
$T=1$ Gogny pairing is included in the RHB calculations of ground
states in the middle panels. The panels on the right-hand side 
display the results
obtained by fully-consistent PN-RQRPA calculations, including the 
$T=1$ proton-neutron pairing channel.}
\label{figE}
\end{figure}

\begin{figure} 
\caption{Gamow-Teller 
strength distributions for 
$^{48}$Ca, $^{90}$Zr and $^{208}$Pb. PN-RQRPA results are 
shown in comparison with experimental data (thick arrows) for the 
GTR excitation energies in $^{48}$Ca~\protect\cite{And.85}, 
$^{90}$Zr~\protect\cite{Bai.80,Wak.97}, 
and $^{208}$Pb~\protect\cite{Hor.80,Aki.95, Kra.01}.}
\label{figF}
\end{figure}

\begin{figure} 
\caption{The running sum of the GTR strength for $^{208}$Pb. 
The dashed line corresponds to a PN-RRPA calculation with 
only positive-energy $ph$ configurations. For the calculation
denoted by the solid line the RRPA space contains 
configurations formed from occupied 
states in the Fermi sea and empty negative-energy 
states in the Dirac sea. The
total sum of the GT strength is compared to the model 
independent Ikeda sum rule
(dotted line).}
\label{figG}
\end{figure}

\begin{figure}
\caption{
The PN-RQRPA strength distribution of discrete $J^\pi=1^+$
GT$^{-}$ states.  
The panel on the right-hand side contains the positive energy
$\pi p-\nu h$ strength.
The left panel displays the negative energy spectrum 
built from $\pi\alpha-\nu h$ transitions.}
\label{figH}
\end{figure}

\begin{figure}
\caption{The Gamow-Teller 
strength distribution in $^{118}$Sn, calculated for different 
values of the strength parameter of the $T=0$ pairing 
interaction (\protect\ref{pn-pair}). The insert displays 
the corresponding calculated centroids of 
the direct spin-flip GT strenght.}
\label{figI}
\end{figure}

\begin{figure}
\caption{ RHB plus proton-neutron RQRPA results for the energy spacings
between the Gamow-Teller 
resonances and the respective isobaric analog resonances for the 
sequence of even-even $^{112 - 124}$Sn target nuclei. 
The experimental data are from Ref.~\protect\cite{Pham.95}.}
\label{figJ}
\end{figure}

\begin{figure}
\caption{The proton-neutron RQRPA and experimental \protect\cite{Pham.95}
differences between the excitation energies
of the GTR and IAR, as a function of the calculated 
differences between the rms radii of the neutron and proton 
density distributions of even-even Sn isotopes (upper panel). 
In the lower panel the calculated differences $r_n - r_p$
are compared with experimental data \protect\cite{Kra.99}.}
\label{figK}
\end{figure}

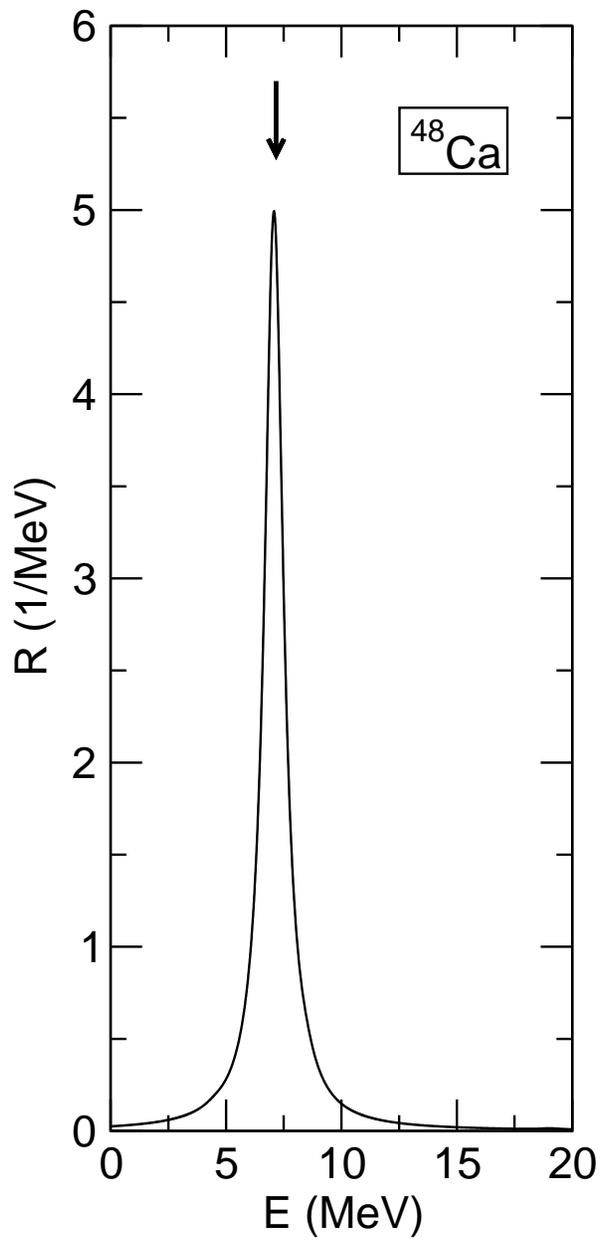
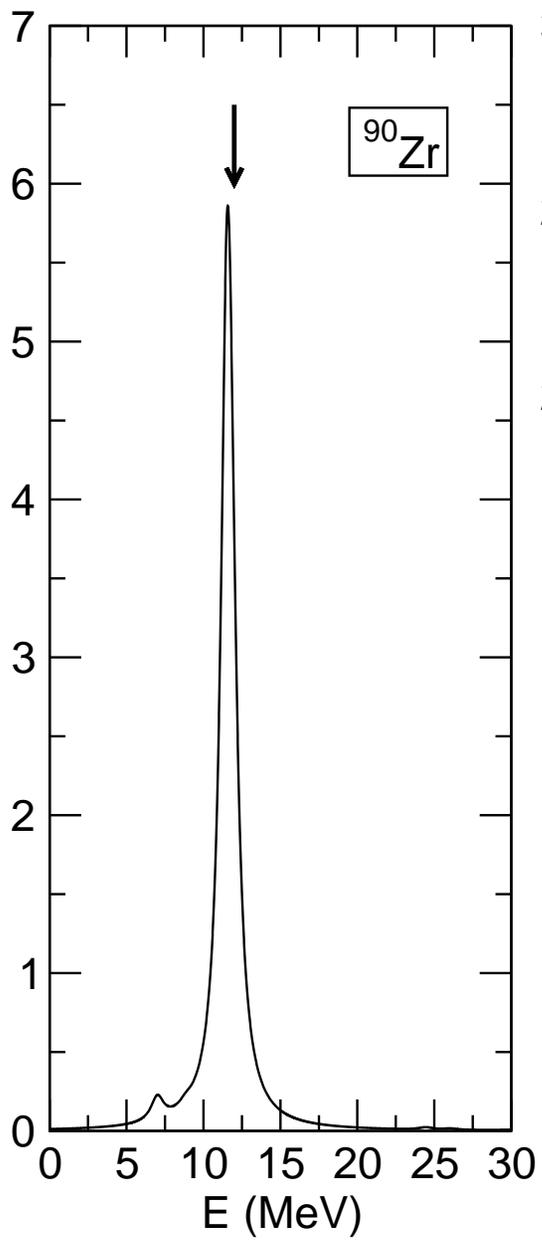
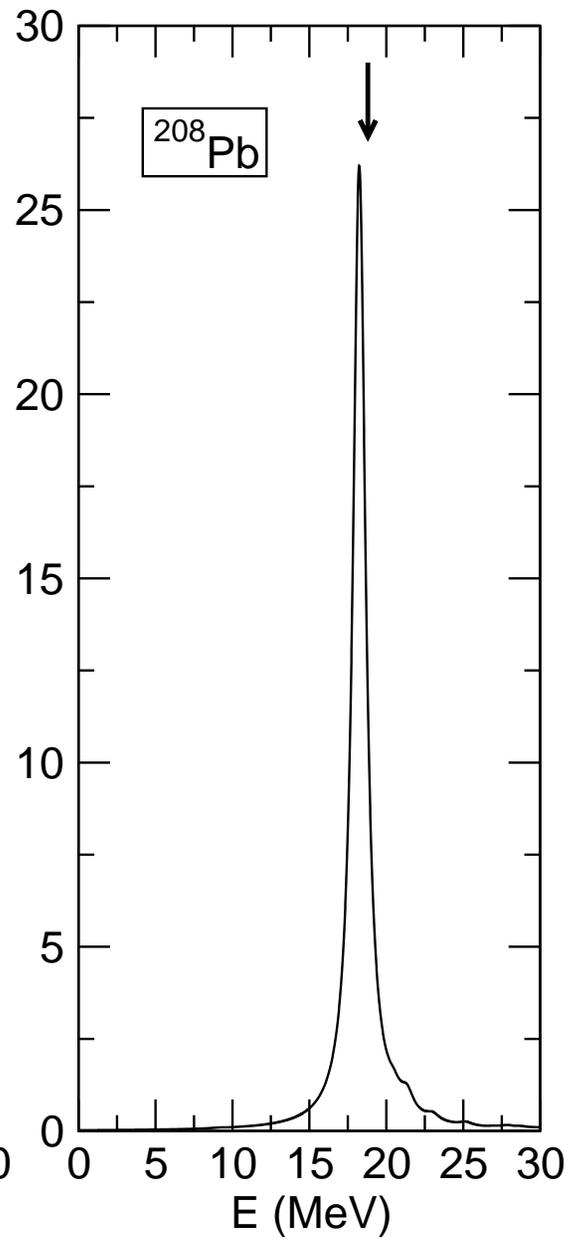
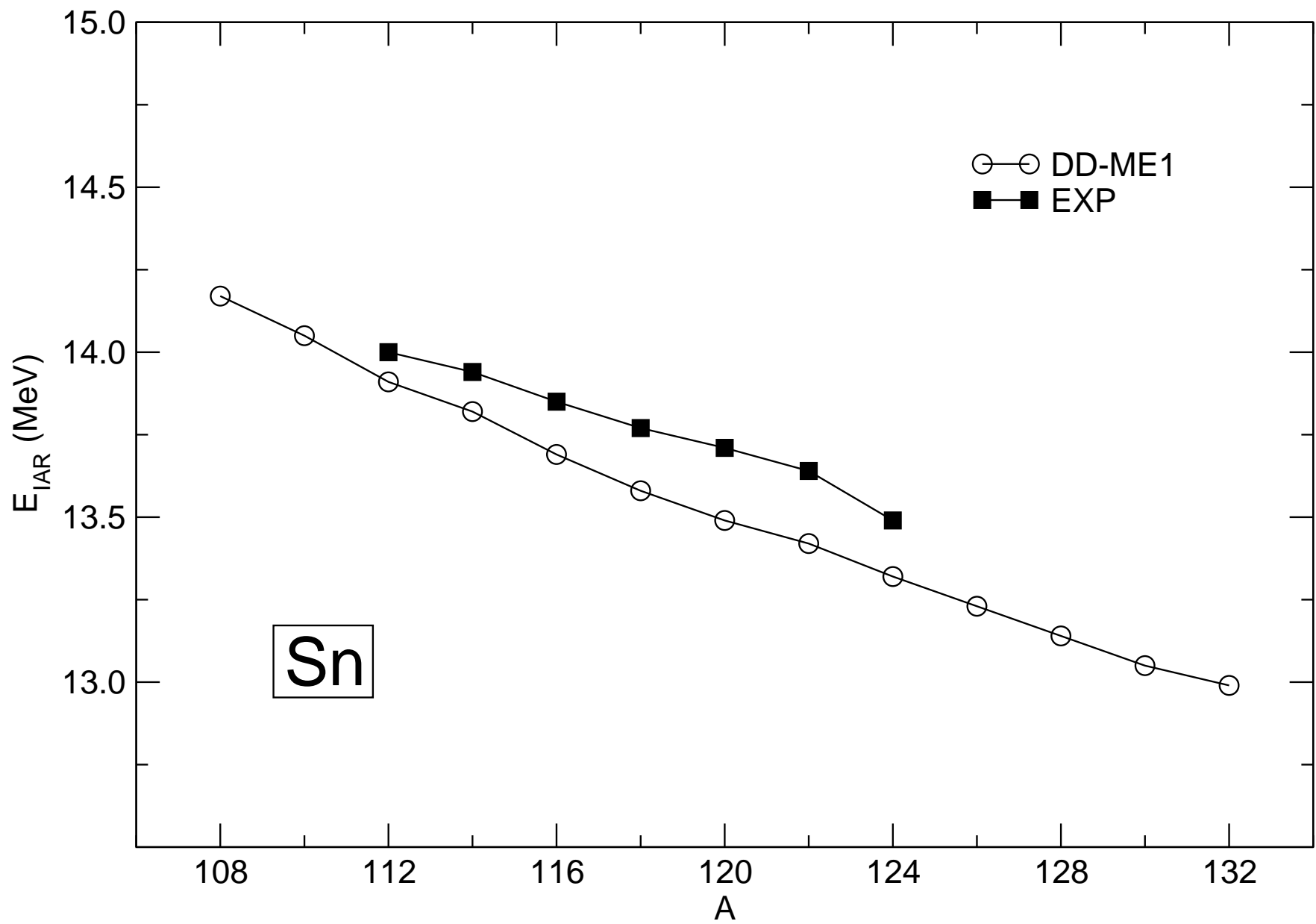
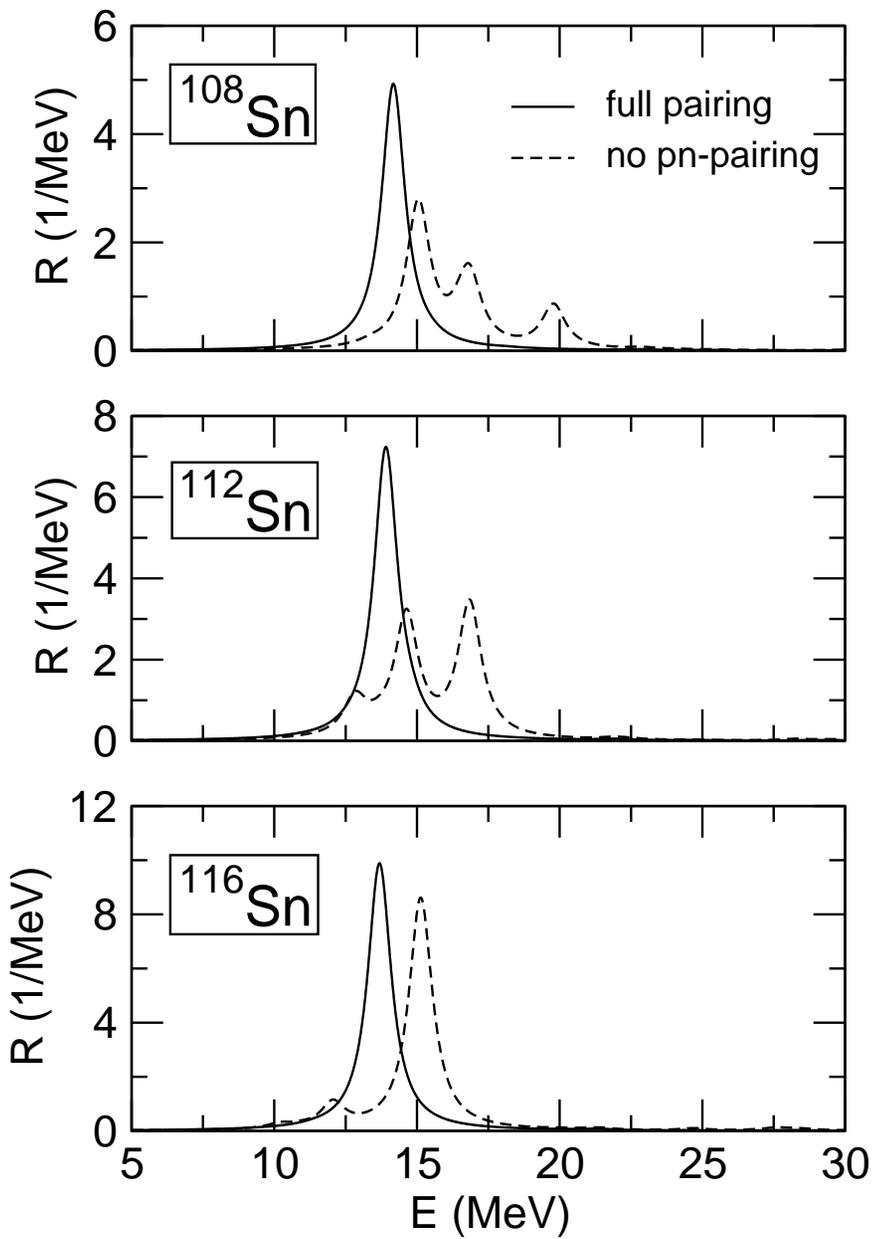
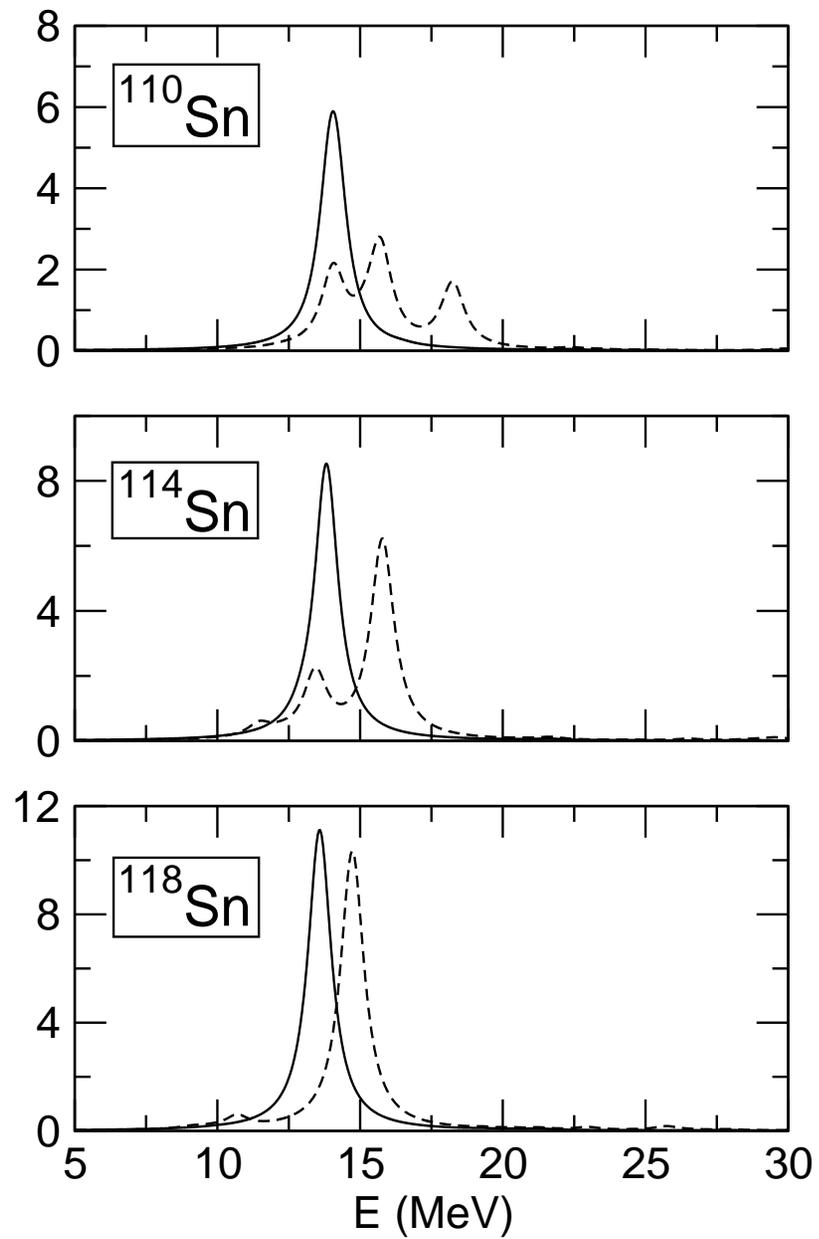
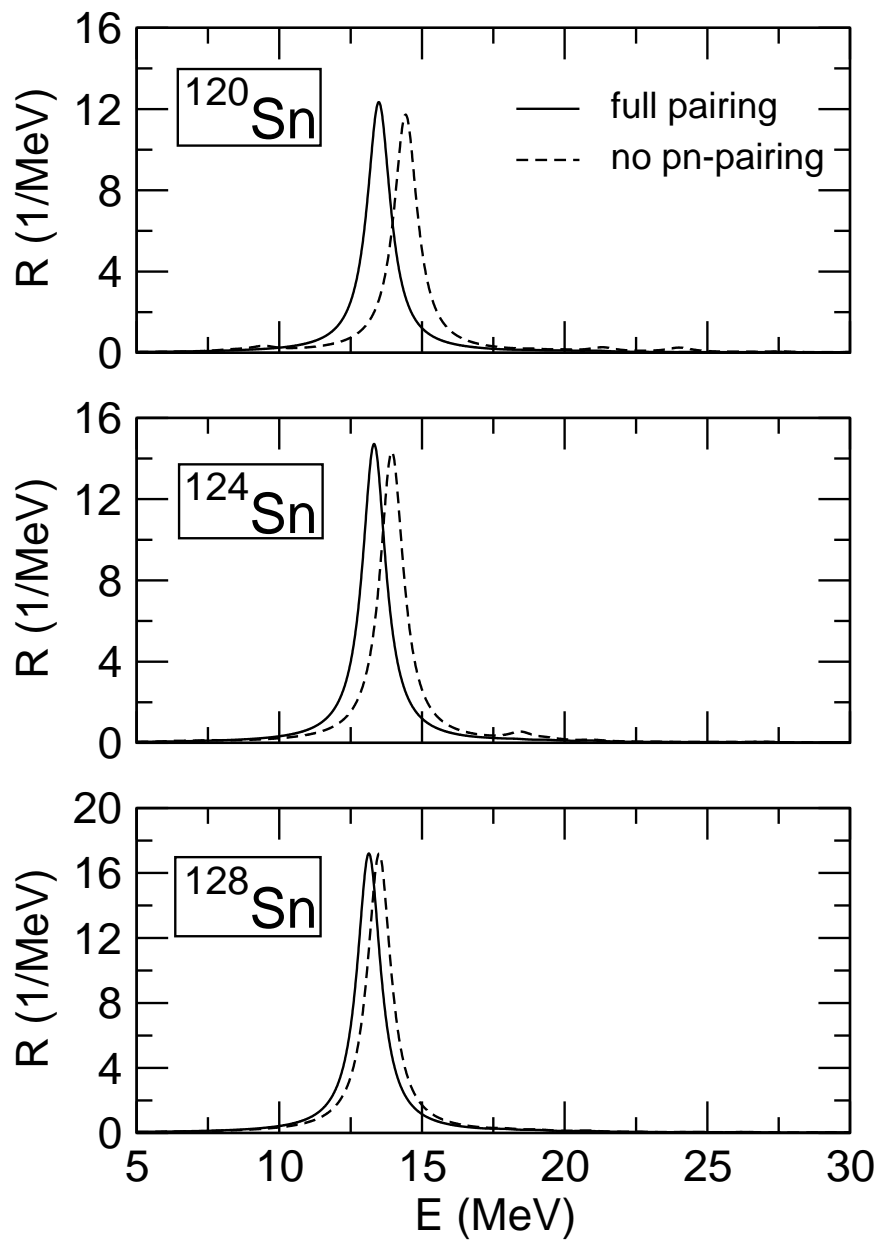
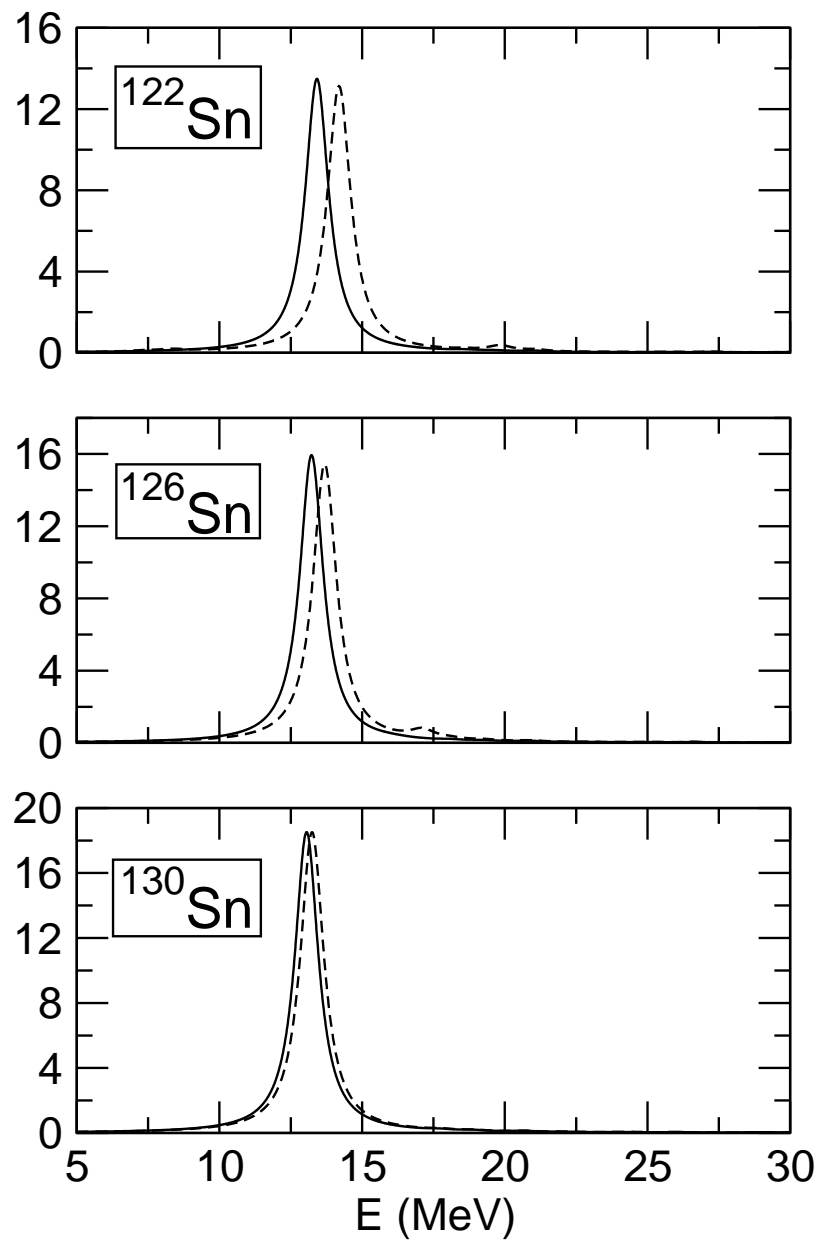
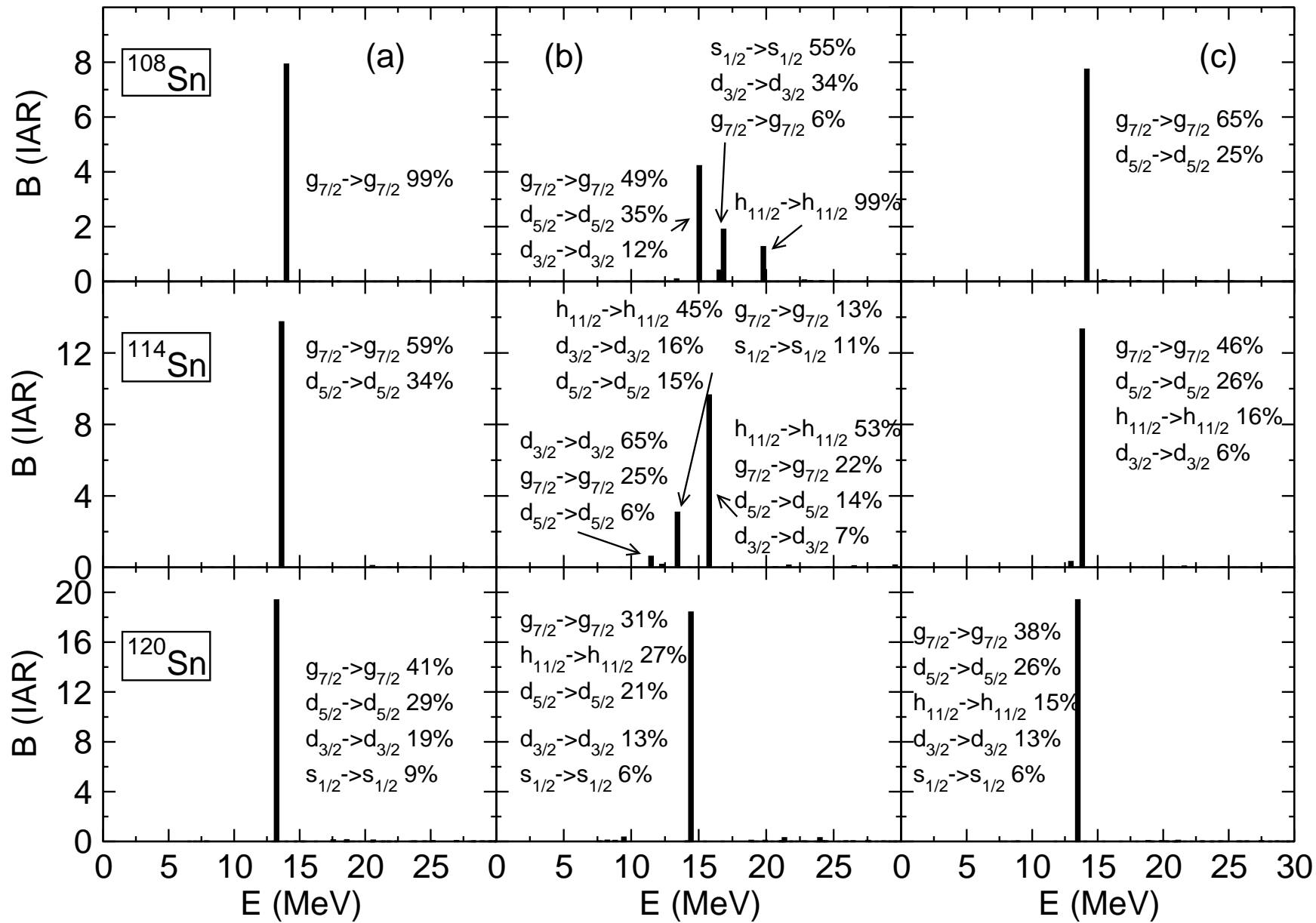
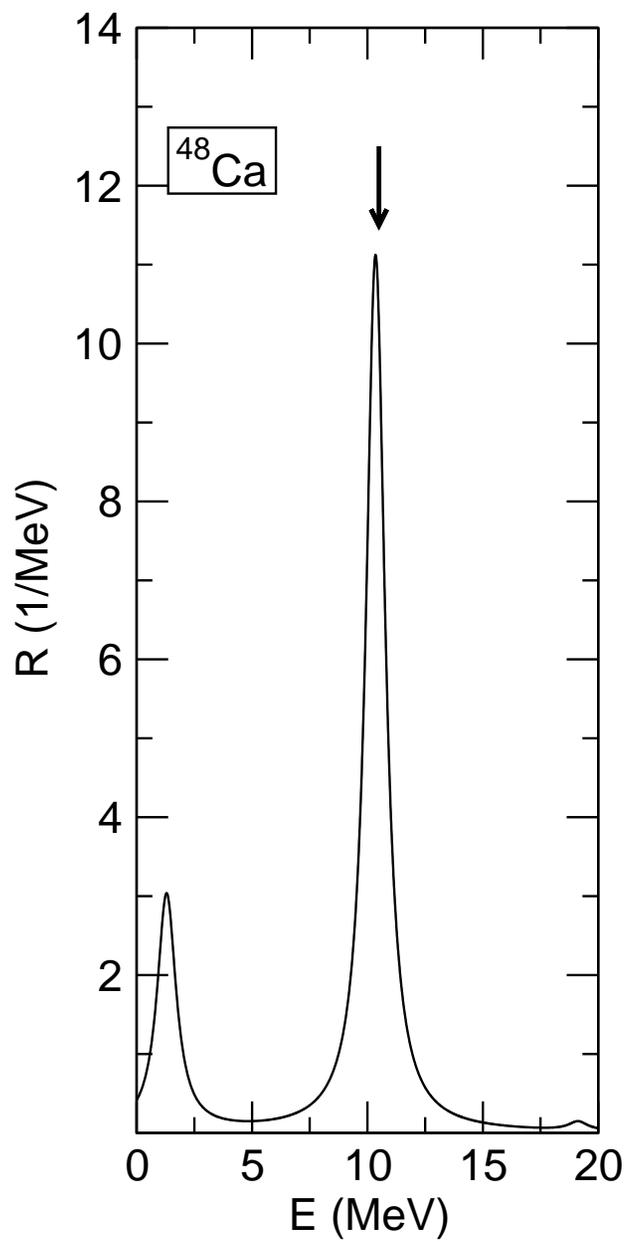
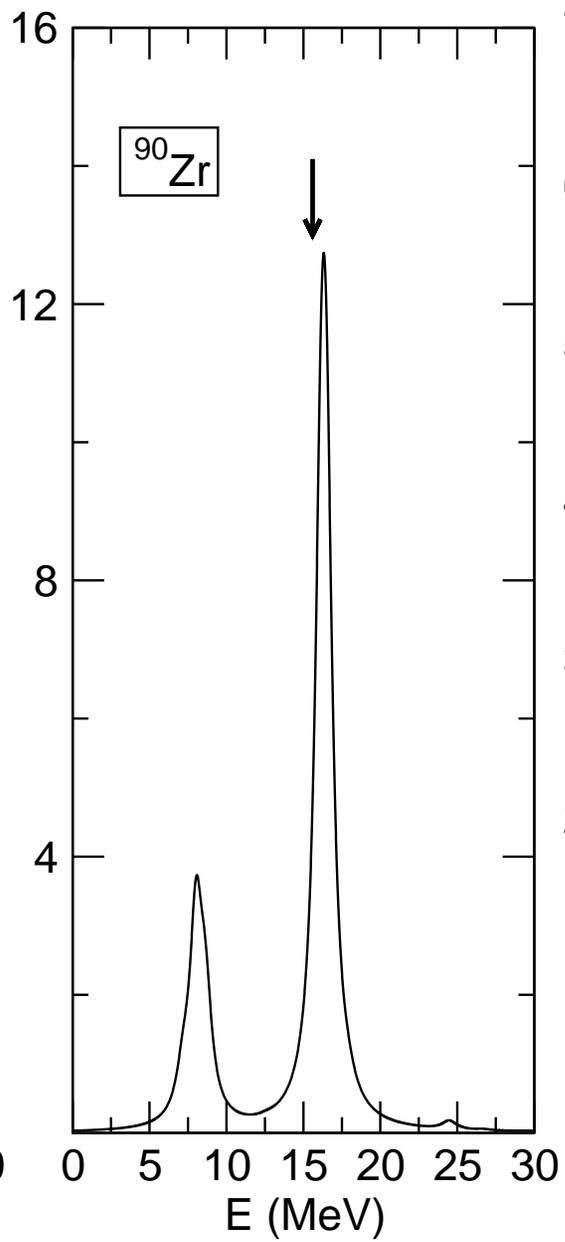
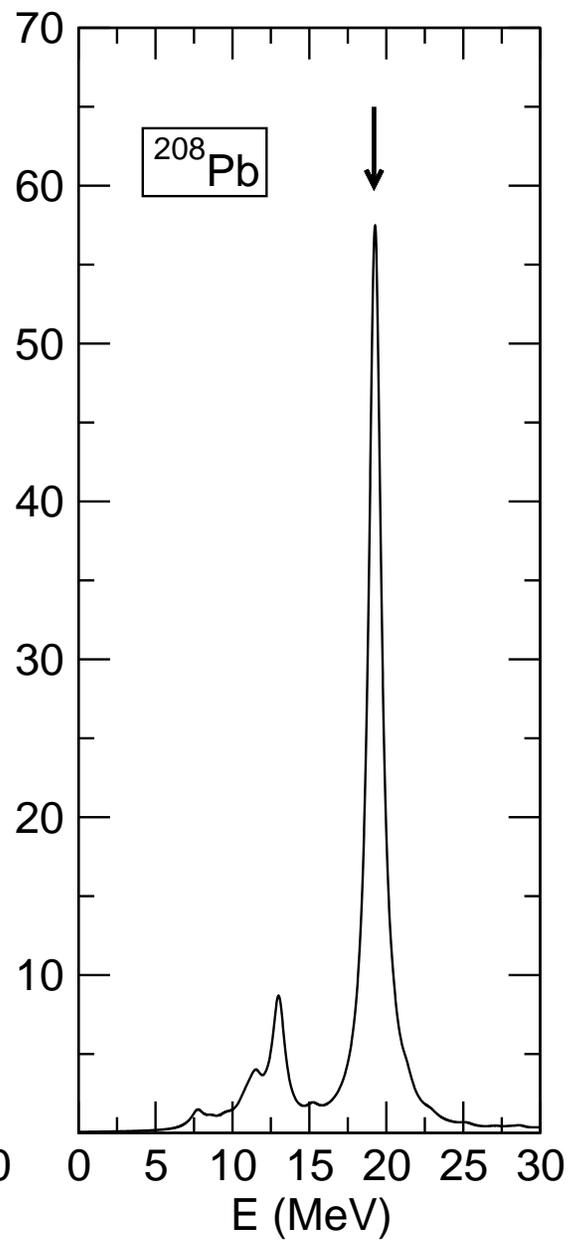
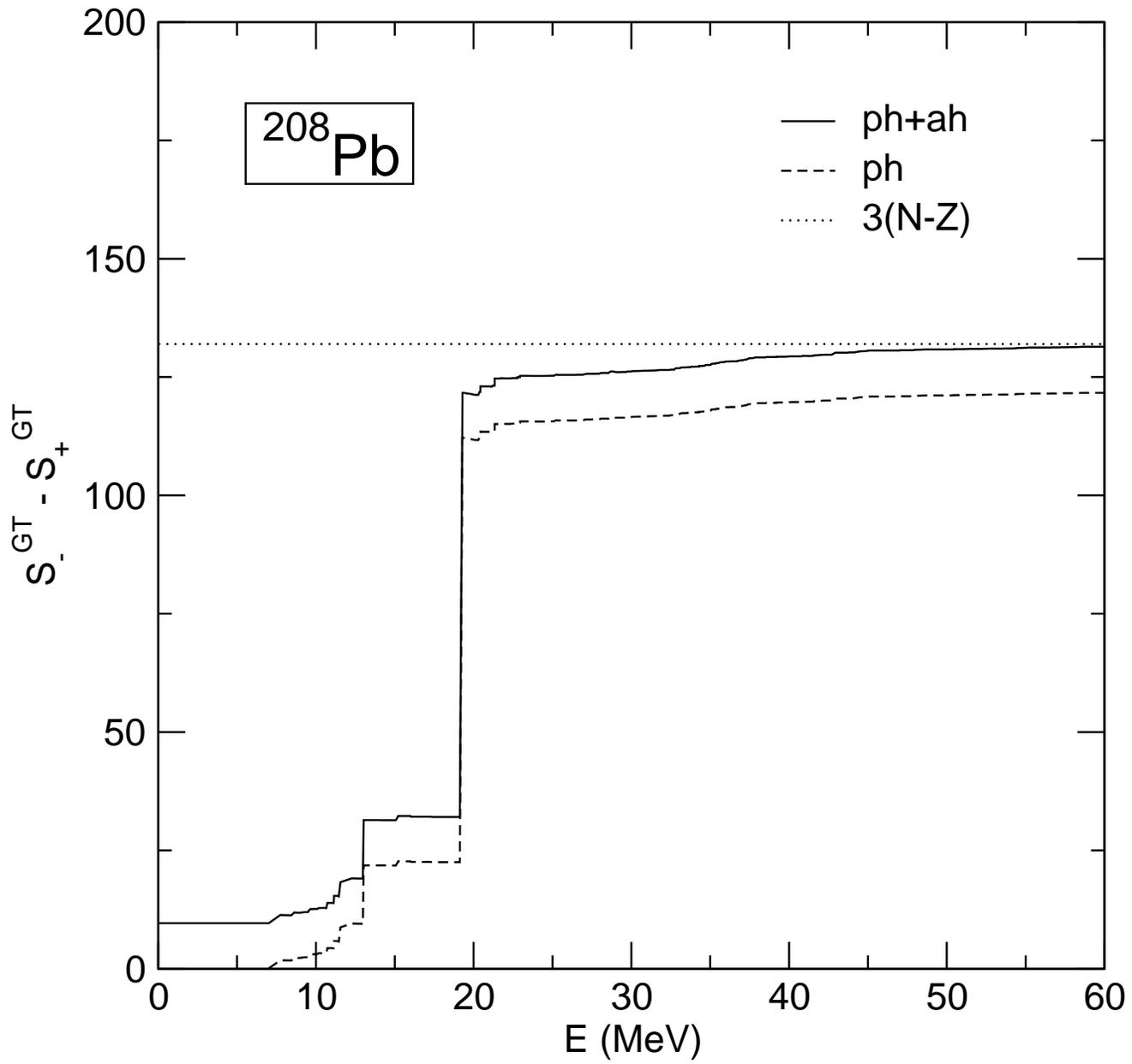
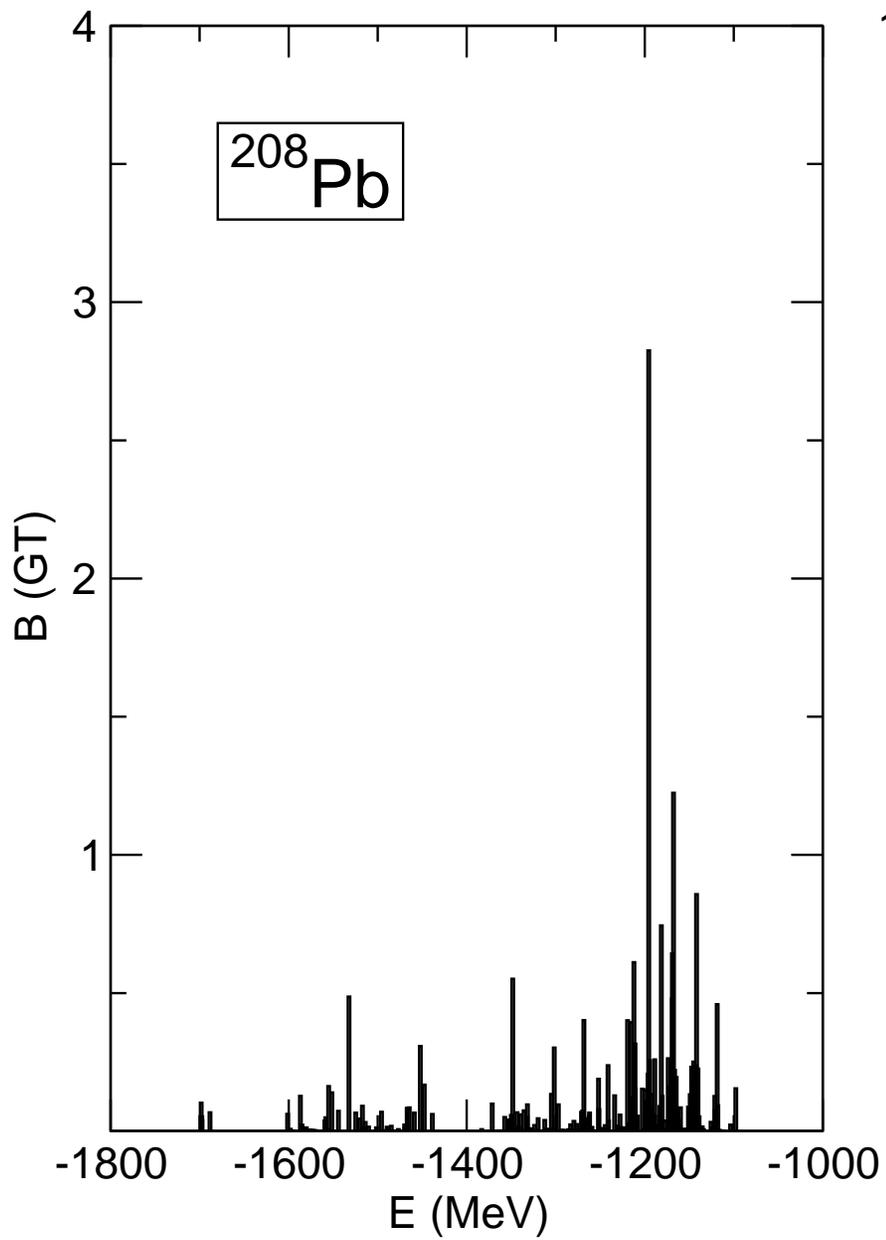
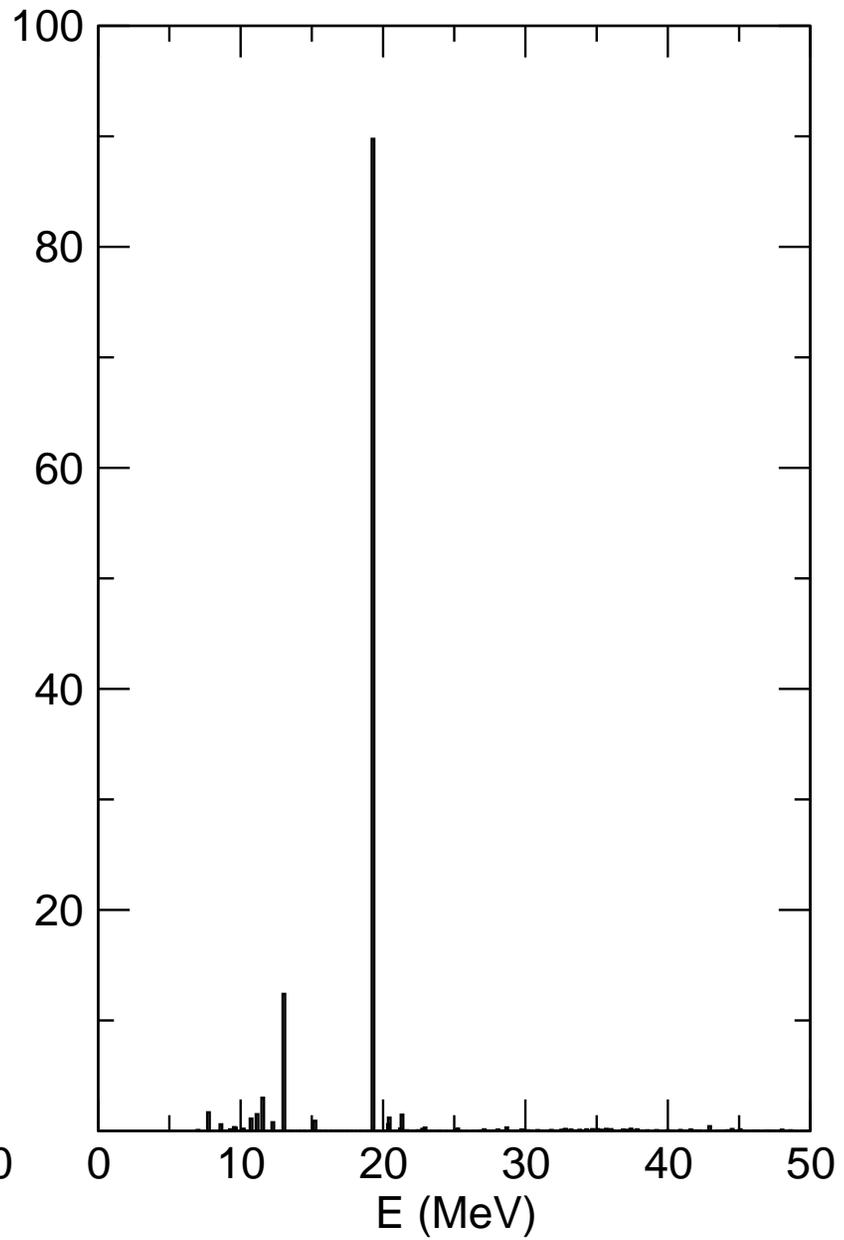
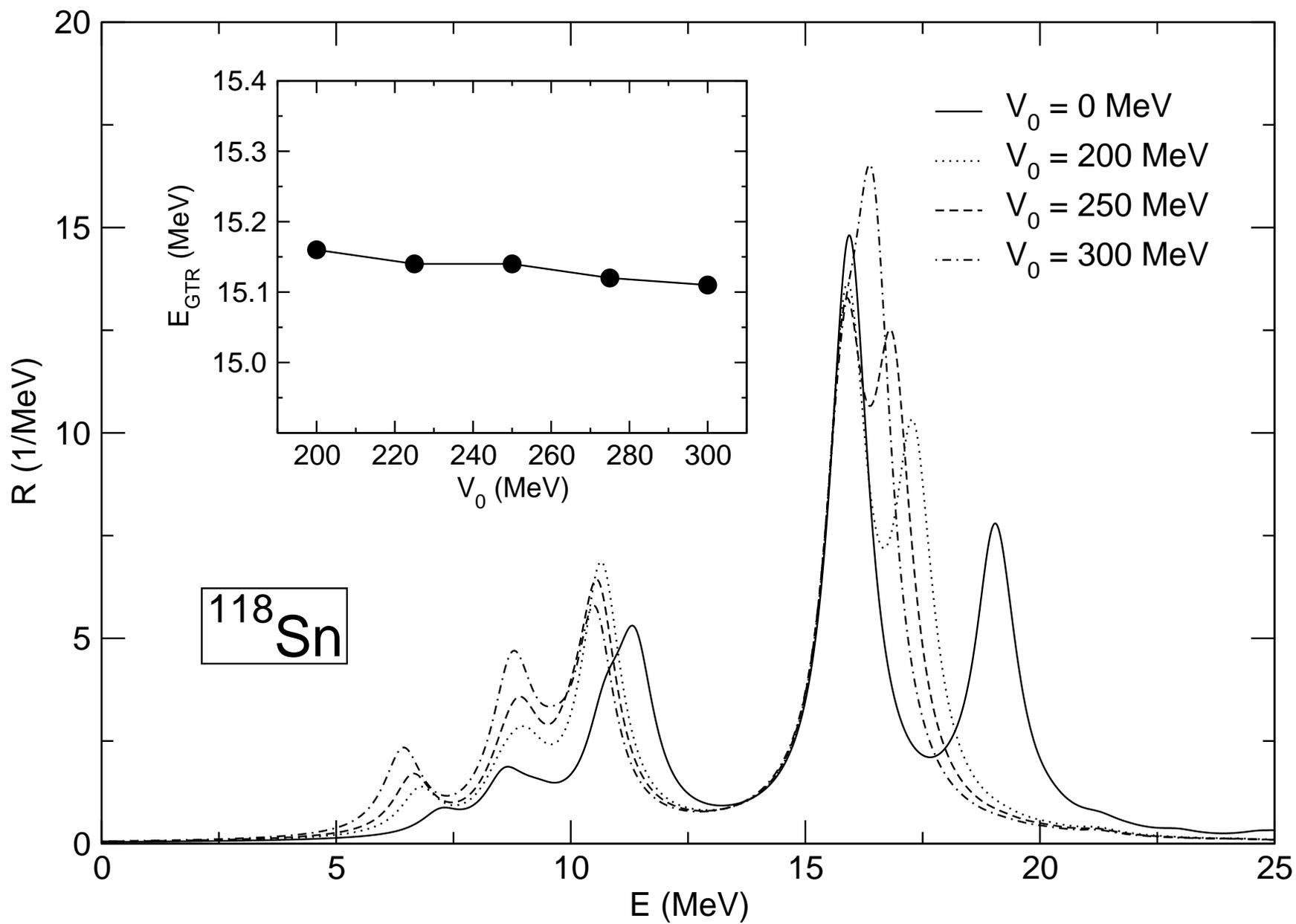
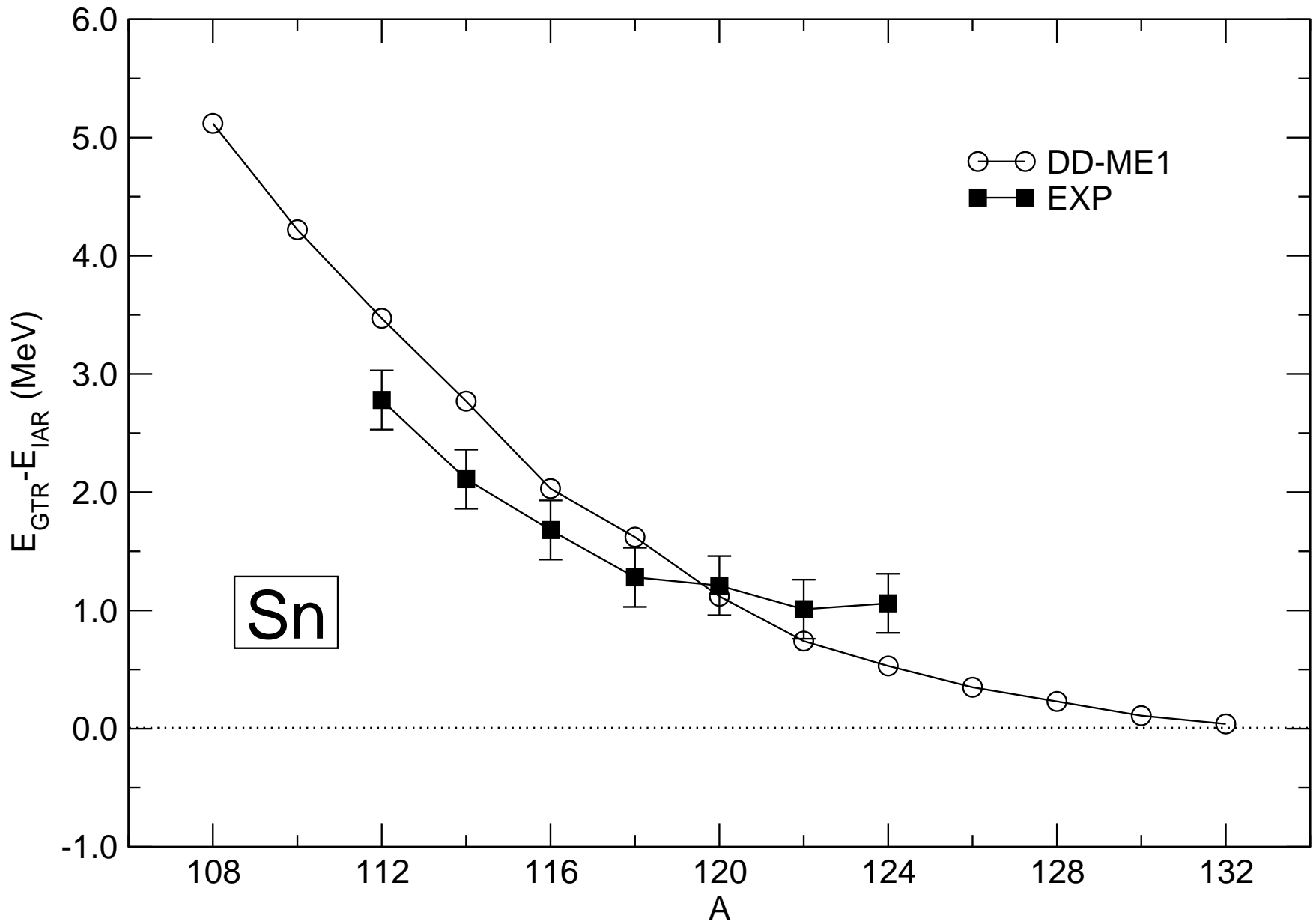
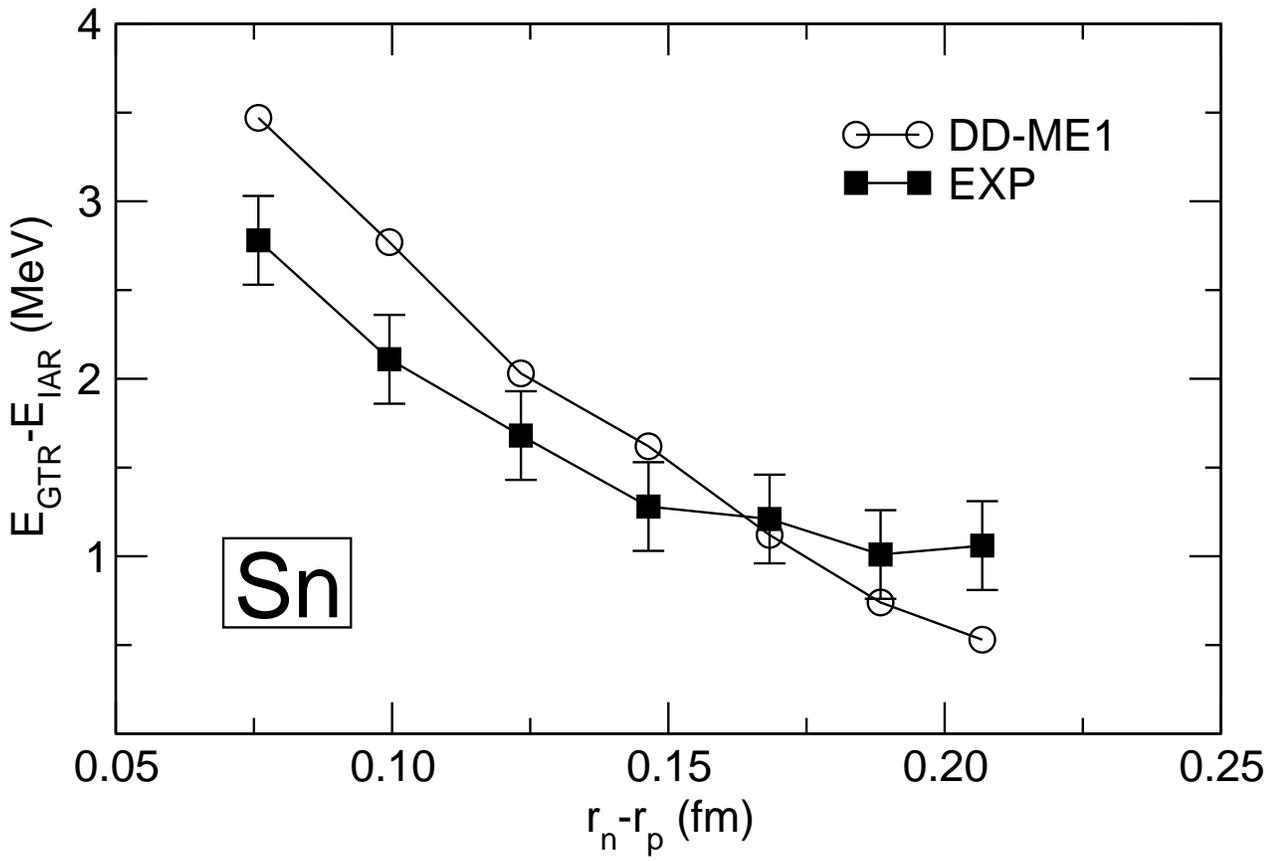
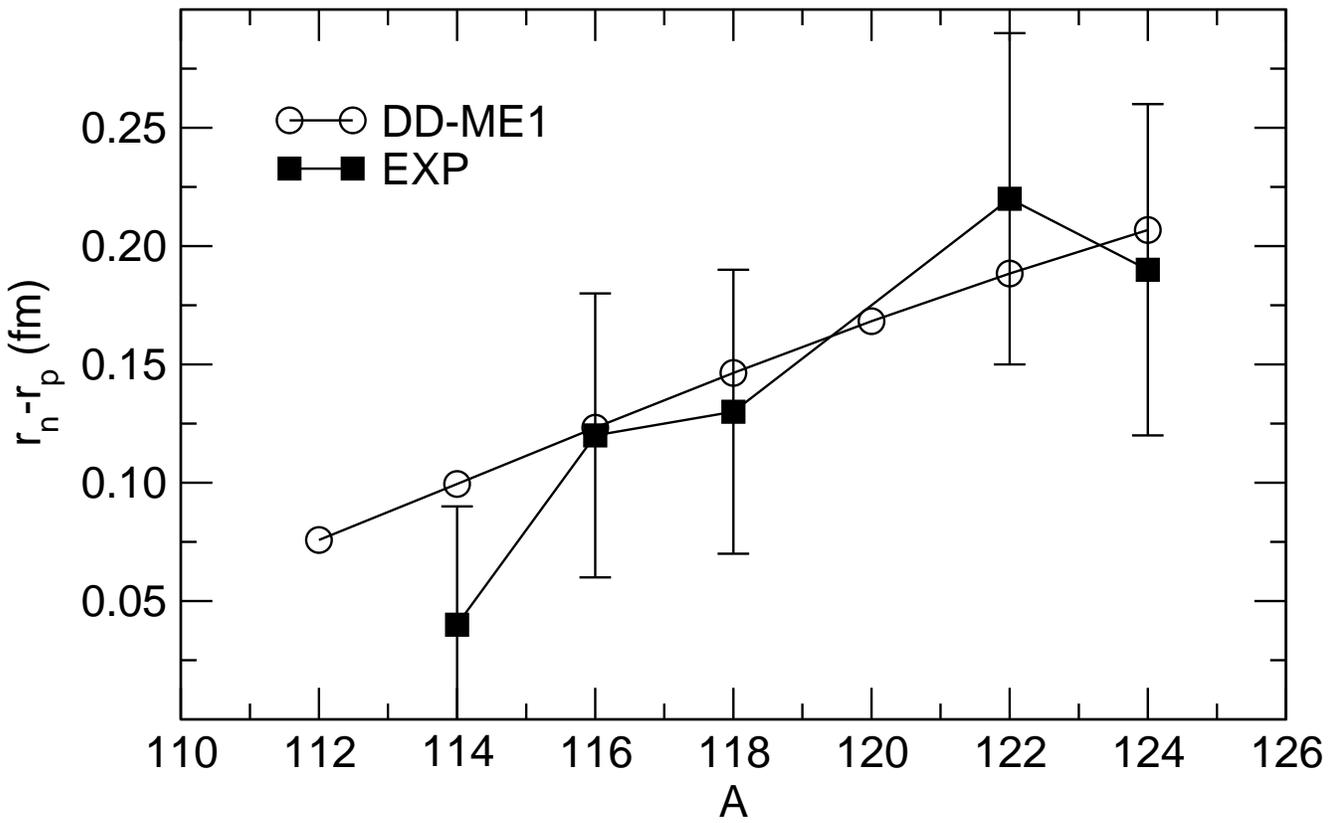
\end{document}